\newif\ifconfver
\confverfalse      
\confvertrue        

\ifconfver
    \documentclass[10pt,twocolumn,twoside]{IEEEtran}
\else
    \documentclass[11pt, draft, onecolumn]{IEEEtran}
\fi

\usepackage{amsmath,epsfig,amssymb,amsthm,subfigure}
\usepackage{graphicx}
\usepackage{psfrag}
\usepackage{cite}
\usepackage[usenames,dvipsnames]{color}
\usepackage{float}

\newtheorem{Prop}{\bf Proposition}

\newtheorem{remark}{\bf Remark}

\newcommand\zerob{\ensuremath{{\bf 0}}}

\newcommand\Hb{\ensuremath{{\bf H}}}

\newcommand\Ab{\ensuremath{{\bf A}}}

\newcommand\Gb{\ensuremath{{\bf G}}}

\newcommand\E{\ensuremath{{\rm E}}}

\newcommand\diag{\ensuremath{{\rm diag}}}

\def\ie{{\it i.e., }}

\def\iid{{\it i.i.d. }}

\def\Nt{{N_t}}
\def\NL{{N_L}}
\def\NU{{N_U}}
\def\cE{\mathcal{E}}
\def\cP{\mathcal{P}}
\def\bH{{\mathbf{H}}}     
\def\sH2{{\sigma_h^2}}
\def\bHd{{\mathbf{H}_d}}
\def\bHu{{\mathbf{H}_u}}
\def\bG{{\mathbf{G}}}

\def\hbHd{{\mathbf{\widehat{H}}_d}}
\def\hbHdt{{\mathbf{\widehat{H}}_{d,t}}}
\def\hbHu{{\mathbf{\widehat{H}}_u}}

\def\sHd2{\sigma_{h_d}^2}
\def\sHu2{\sigma_{h_u}^2}
\def\sG2{\sigma_{g}^2}
\def\bbC{\mathbb{C}}
\def\bI{\mathbf{I}}

\def\diag{{\rm diag}}
\def\calC{{\mathcal C}}
\def\calN{{\mathcal N}}
\def\bR{{\mathbf{R}}}
\def\bh{{\mathbf{h}}}
\def\by{{\mathbf{y}}}

\def\bA{{\mathbf{A}}}
\def\sa2{{\sigma_a^2}}

\def\bNHdt{{\mathbf{K}_{\widehat{\bH}_{d,t}}}}

\def\bNHdHt{{\mathbf{K}^H_{\widehat{\bH}_{d,t}}}}

\def\bV{{\mathbf{V}}}
\def\sv2{{\sigma_v^2}}

\def\bX{{\mathbf{X}}}
\def\bF{{\mathbf{F}}}
\def\bC{{\mathbf{C}}}
\def\bY{\mathbf{Y}}
\def\bW{\mathbf{W}}
\def\sw2{\sigma_w^2}
\def\bWt{{\mathbf{\widetilde{W}}}}
\def\swt2{\sigma_{\tilde{w}}^2}
\def\bPave{P_{ave}}  
\def\bPL{\bar{P}_L}  
\def\bPt{\bar{P}_t}  
\def\bEave{\bar{\mathcal{E}}_{tot}}
\def\bEL{\bar{\mathcal{E}}_L}
\def\bEt{\bar{\mathcal{E}}_t}
\def\tgam{{\tilde{\gamma}}}   

\psfull
\begin{document}

\markboth{Accepted by IEEE TRANSACTIONS ON Signal Processing, Jan.
2013}{Accepted by IEEE TRANSACTIONS ON Signal Processing, Jan. 2013}

\title{Two-Way Training for Discriminatory Channel Estimation in Wireless MIMO Systems}
\author{\vspace{0.5cm}
\IEEEauthorblockN{Chao-Wei Huang, Tsung-Hui Chang, Xiangyun Zhou, and Y.-W. Peter Hong}
\thanks{Copyright (c) 2012 IEEE. Personal use of this material is permitted. However, permission to use this material for any other purposes must be obtained from the IEEE by sending a request to pubs-permissions@ieee.org.}
\thanks{Chao-Wei Huang and Y.-W. Peter Hong are with Institute of Communications Engineering,
National Tsing Hua University, Hsinchu, Taiwan 30013, R.O.C. E-mail: cwhuang@erdos.ee.nthu.edu.tw,
ywhong@ee.nthu.edu.tw. Tsung-Hui Chang is with Department of Electronic and Computer Engineering,
National Taiwan University of Science and Technology, Taipei 106, Taiwan (R.O.C). E-mail:
tsunghui.chang@ieee.org. Xiangyun Zhou is with Research School of Engineering, The Australian
National University, Canberra, ACT 0200, Australia. E-mail: xiangyun.zhou@anu.edu.au. Y.-W. Peter
Hong is the corresponding author.}\thanks{This work was supported in part by the National
Science Council, Taiwan, under grant NSC 100-2628-E-007-025-MY3 and Grant NSC
101-2218-E-011-043, and in part by the Australian
Research Council's Discovery Projects funding scheme (project no. DP110102548).}}

\maketitle


\vspace{-0.0cm}
{
\begin{abstract}
This work examines the use of two-way training to efficiently discriminate the channel estimation performances at a legitimate receiver (LR) and an unauthorized receiver (UR) in a multiple-input multiple-output (MIMO) wireless system. This work improves upon the original discriminatory channel
estimation (DCE) scheme proposed by Chang {\it et al} where multiple stages of feedback and retraining were used. While most studies on physical layer secrecy are under the information-theoretic framework and focus directly on the data transmission phase, studies on DCE
focus on the training phase and aim to provide a practical signal processing technique to discriminate between the channel estimation performances (and, thus, the effective received signal qualities) at LR and UR. A key feature of DCE designs is the insertion of artificial noise (AN) in the training signal to degrade the channel estimation performance at UR. To do so, AN must be placed in a carefully chosen subspace based on the transmitter's knowledge of LR's channel in order to minimize its effect on LR. In this paper, we adopt the idea of two-way training that allows both the transmitter and LR to
send training signals to facilitate channel estimation at both ends. Both reciprocal and non-reciprocal channels are considered and a two-way DCE scheme is proposed for each scenario. {For mathematical tractability, we assume that all terminals employ the linear minimum mean
square error criterion for channel estimation. Based on the mean square error (MSE) of the channel estimates at all terminals,} we formulate and solve an optimization problem where the optimal power allocation between the training signal and AN is found by minimizing the MSE of LR's channel
estimate subject to a constraint on the MSE achievable at UR. Numerical results show that the proposed DCE schemes can effectively discriminate between the channel estimation and hence the data detection performances at LR and UR.
 \ifconfver
\else
\\\\
\noindent {\bfseries EDICS}: SPC-CEST, MSP-CEST, SPC-APPL, MSP-APPL  \fi
\end{abstract}
}





\section{Introduction}
Due to the broadcast nature of the wireless medium, communication between wireless terminals is
often susceptible to potential eavesdropping by unauthorized receivers. Therefore, as wireless
technology becomes more prevalent, the need for discriminating between the signal reception
performance at a legitimate receiver (LR) and that at an unauthorized receiver (UR) has increased.
Motivated by this demand, the concept of physical layer secrecy have been studied extensively in
recent years and methods that utilize properties of the wireless channels to achieve the desired
performance discrimination have been proposed. From an information-theoretic viewpoint, the
difference in the channel condition at different receivers can be exploited to ensure a non-zero
communication rate between the transmitter and LR under the perfect secrecy constraint
\cite{secrecy.capacity,Khisti,Oggier,Goel}, where the notion of perfect secrecy means that no UR is
able to infer any information from the received signal broadcasted by the transmitter.
From a signal processing perspective, {artificial-noise-aided multiuser beamforming schemes
\cite{Swindlehurst,Swindlehurst2,Liao} and space-time coding schemes \cite{Fakoorian2011}} can be
adopted to enhance signal reception at LR while limiting the received signal quality at UR.

\subsection{Motivation and Related Work}


Most studies in the literature on physical layer secrecy, \emph{e.g.},
\cite{secrecy.capacity,Khisti,Oggier,Goel,Swindlehurst,Swindlehurst2,Liao,Fakoorian2011}, focus on
the design of the data transmission phase while often assuming the availability of perfect channel
state information. {In practice, channel knowledge is typically obtained through training and
channel estimation, and its quality can have a significant impact on the receiver performance
\cite{YooGoldsmith06,Hassibi_howmuchtraining}.
Intuitively, if UR has a poorer channel estimation performance than LR, then
UR would have a higher detection error probability when overhearing the transmission of secret messages.} 
Motivated by this observation, the authors in \cite{ChangChiangHongChi_TSP2010} proposed a novel
training scheme, called {\it discriminatory channel estimation} (DCE), that can provide a better
channel estimation performance for LR compared to that for UR. Different from the
information-theoretic works \cite{secrecy.capacity,Khisti,Oggier}, the DCE scheme in
\cite{ChangChiangHongChi_TSP2010} takes a more practical viewpoint and is a signal processing
technique for discriminating between the channel conditions of LR and UR.

{The key feature of the DCE scheme \cite{ChangChiangHongChi_TSP2010} is the insertion of artificial
noise (AN) in the training signal. The AN is carefully placed in a subspace so that it jams UR
while the interference caused to LR is minimized. To this end, the transmitter requires knowledge
of LR's channel.}
%
In the original DCE scheme proposed in \cite{ChangChiangHongChi_TSP2010}, this was achieved by
using a preliminary training stage followed by multiple stages of feedback-and-retraining.
Specifically, in the preliminary training stage,  a pure pilot signal is first sent by the
transmitter to allow for a rough channel estimate at LR. Then, LR feeds back this channel estimate
to the transmitter, who then sends a new pilot signal inserted with AN to disrupt the channel
estimation at UR. The pilot signal power used in the preliminary training stage must be relatively
small since, otherwise, UR is also able to obtain a good channel estimate. In addition, the AN signal used
in the retraining stage must be placed in the null-space of LR's estimated channel to minimize its
effect on LR. Through each stage of feedback and retraining, knowledge of LR's channel at the
transmitter is gradually refined, allowing AN to be placed more precisely in the desired signal
subspace and the pilot signal power to be increased without benefiting the channel estimation at
UR. The main drawback of this multi-stage feedback-and-retraining scheme
\cite{ChangChiangHongChi_TSP2010} is the large training overhead and the high design complexity
required. The cost and the quality of the feedback link may also limit its application in practice,
but was not considered in \cite{ChangChiangHongChi_TSP2010}.

\subsection{Our Approach and Contribution}

The main contribution of this work is to propose new and efficient DCE schemes based on the two-way
training methodology. The idea of two-way training, which has been studied for non-secrecy
applications, \emph{e.g.}, in
{\cite{twoway.SIMO,twoway.MISO,twoway.MIMO,twoway.ECHOMIMO,Dong09,Zhang09}}, allows both the
transmitter and the receiver to send pilot signals in a collaborative manner so that channel
estimation is enabled at both ends. {This is particularly useful in achieving DCE since the reverse
training signal sent by LR in a two-way training scheme will not benefit UR in obtaining any
information about the channel between itself and the transmitter\footnote{However, the revere
training signal enables UR to estimate the channel between itself and LR. This may not be desirable
if LR also has secret messages to transmit; see more discussions in Remark 5.}}. This advantage is
not enjoyed by conventional one-way training schemes since any pilot signal sent by the transmitter
will help UR improve its estimate of the channel between itself and the transmitter. In this work,
two-way DCE schemes  are designed for both reciprocal and non-reciprocal channel models. The former
is a reasonable model for time-division duplex (TDD) systems whereas the latter is often used to
model frequency-division duplex (FDD) systems. For reciprocal channels, the proposed two-way DCE
scheme requires only two stages of training, that is, a reverse training stage and a forward
training stage. For non-reciprocal channels, only an additional round-trip training stage is
needed, in which the transmitter first broadcasts a random signal only known to itself and LR
echoes the signal back using an amplify-and-forward strategy. In both cases, AN is inserted into
the pilot signal in the (final) forward training stage to achieve the desired DCE performance.
Compared to the multi-stage feedback-and-retraining scheme proposed
in~\cite{ChangChiangHongChi_TSP2010}, the newly proposed two-way training schemes drastically
reduce the overall training overhead.

The proposed two-way DCE schemes can conceptually be derived under any channel estimation criterion
at the three terminals. {For tractability and for gaining useful insights, in this paper, we assume
that all terminals employ the linear minimum mean square error (LMMSE) channel estimator, and
derive the resulting mean square error (MSE) of the channel estimates obtained at both LR and UR.}
These analysis results are then used to compute the optimal power allocation between the pilot
signals and AN across different training stages. {This is obtained by solving an optimization
problem that aims to minimize the MSE of LR's channel estimate subject to a constraint on the MSE
of UR's channel estimate and individual training energy constraints at the transmitter and LR. For reciprocal channels, we show that the optimal transmit powers of reverse
training, forward training, and AN have simple close-form expressions.} For non-reciprocal channels, the power allocation problem cannot
be easily solved, but an approximate solution can be obtained by employing the monomial
approximation and the condensation method {\cite{Tutorial_GP2,Tutorial_GP}} often used in the field
of geometric programming (GP). Numerical results are provided to demonstrate the effectiveness of
the proposed schemes.

The remainder of this paper is organized as follows. {In Section~\ref{ch3.Recip}, the system model
and the proposed two-way DCE scheme for reciprocal channels are presented. In
Section~\ref{ch4.NonRecip}, the two-way DCE scheme is extended to non-reciprocal channels.}
Numerical results are provided in Section~\ref{ch7.simulation} and, finally, the conclusion is
given in Section~\ref{ch8.conclusion}.

{\bfseries Notations}: Upper-case and lower-case boldfaced letters are used for matrices, \emph{e.g.}, $\mathbf{X}$, and vectors, \emph{e.g.}, $\mathbf{x}$, respectively.
Moreover, $\mathbf{X}^T$, $\mathbf{X}^*$ and $\mathbf{X}^H$ denote the transpose, the complex
conjugate and the Hermitian of the matrix $\mathbf{X}$, respectively. Let $\mathbf{0}_{M\times N}$
be the $M$-by-$N$ zero matrix and let $\bI_M$ be the $M$-by-$M$ identity matrix. $\mbox{Tr}(\cdot)$ denotes
the trace of a square matrix, $\|\cdot\|$ denotes the Frobenius norm, and $\mbox{vec}(\cdot)$ is the operator which stacks the columns of
a matrix into a vector. The symbol $\otimes$ denotes the Kronecker matrix product, $\mathbb{E}\{\cdot\}$
denotes the expectation operator and $\diag(a_1,\dots,a_N)$ is the $N\times N$ diagonal matrix with
diagonal elements $a_1,\dots,a_N$.

\section{Two-Way DCE Design for Reciprocal Channels}\label{ch3.Recip}

\subsection{System Model} \label{ch2.system}
\begin{figure}[t]
\centering {\includegraphics[scale=0.6]{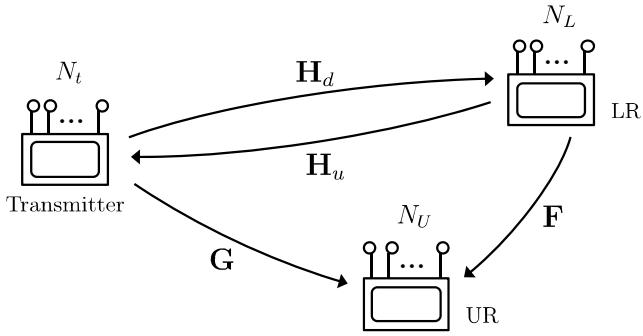}} \vspace{-0.0cm}\caption{A wireless MIMO system
consisting of a transmitter, a legitimate receiver (LR) and an unauthorized receiver (UR).}
\label{fig.ch2.sysmodel}\vspace{-0.0cm}
\end{figure}

Consider a wireless MIMO system {that consists} of a transmitter, a legitimate receiver (LR), and
an unauthorized receiver (UR)\footnote{For ease of presentation, we will focus on the scenario with
one LR and one UR throughout the paper. The presented methods, nevertheless, can be easily extended
to the scenario with multiple LRs and URs, following the same spirit as in
\cite{ChangChiangHongChi_TSP2010}.}, which are equipped with $\Nt$, $N_L$ and $N_U$ antennas,
respectively, as shown in Fig.~\ref{fig.ch2.sysmodel}. We assume that $\Nt>\NL$. Moreover, the
channel fading coefficients are assumed to remain constant during each transmission block, which
consists of a training phase and a data transmission phase. In this work, we focus on the training
phase and aim to discriminate the channel estimation performances at LR and UR. Let the downlink
channel matrix from the transmitter to LR be denoted by $\bHd \in \bbC^{\Nt \times \NL}$. The
entries of $\bHd$ are assumed to be independent and identically distributed ({\it i.i.d.}) complex
Gaussian random variables with zero mean and variance $\sHd2$ (\ie $\calC\calN(0,\sHd2)$).
Similarly, the uplink channel from LR to the transmitter is denoted by $\bHu \in \bbC^{\NL \times
\Nt}$, whose entries are \iid with distribution $\calC\calN(0,\sHu2)$. {The forward channel from
the transmitter to UR is denoted by $\bG \in \bbC^{\Nt \times \NU}$ with entries being \iid
$\calC\calN(0,\sG2)$; while the channel from LR to UR is denoted by $\bF \in \bbC^{\NL \times
\NU}$. In the rest of the paper, we assume that the transmitter, LR and UR are separated enough so
that $\bHd$ ($\bHu$), $\bG$ and $\bF$ are distinct and statistically independent of each other. }

In the case of reciprocal channels, the uplink and downlink channel matrices can be specified by a
single channel matrix $\bH$ such that $\bH\triangleq \bHd=\mathbf{H}^T_u \in \mathbb{C}^{\Nt \times
N_L}$. The variance of each entry can be denoted by $\sH2$, where $\sH2= \sHd2=\sHu2$. With such
channel reciprocity, the transmitter can obtain an estimate of the downlink channel by taking the
transpose of the estimated channel matrix obtained through reverse training, \ie training based on
the pilot signals sent from LR to the transmitter. {In the following subsections, we show the
detailed steps of the proposed two-way DCE scheme and the associated LMMSE channel estimation
performance in reciprocal channels.}

\subsection{Training Scheme and Channel Estimation Performance}
\label{chap2-2}
{\bf Step I (Channel Acquisition at Transmitter via Reverse Training)}:
{The first step of the two-way DCE scheme is to allow the transmitter to obtain a reliable estimate of its downlink
channel to LR without benefiting the channel estimation process at UR. Different from \cite{ChangChiangHongChi_TSP2010}, where the availability of a noiseless feedback
link was considered, our proposed two-way DCE scheme requires the transmitter to estimate the downlink channel by itself through the
exchange of training signals between the transmitter and LR. In the reciprocal channel case, this can be simply achieved by having LR send a reverse training signal to the transmitter.} Specifically, in the reverse training stage, LR first sends a training signal
\begin{equation} \label{equ.training.LR}
\bX_L=\sqrt{\frac{\cP_R\tau_R}{N_L}}\bC_L
\end{equation}
to the transmitter, where $\bC_L\in\bbC^{\tau_R \times \NL}$ is the pilot matrix that satisfies $\bC^H_L\bC_L=\bI_\NL$, $\cP_R$ is the transmission power for reverse training, and $\tau_R$ is the reverse training length. Note that the choice of using an orthonormal
pilot matrix (\ie $\bC^H_L\bC_L=\bI_\NL$) is due to its optimality in minimizing the channel
estimation error in point-to-point channels, as shown in  \cite{Hassibi_howmuchtraining}. In the remainder of this paper, we shall sometimes denote the
reverse training energy by $\cE_R$ while keeping in mind that $${ \cE_R\triangleq \cP_R \tau_R.}$$ The signal received at the transmitter is given by
\begin{equation}\label{equ.rec.tx}
\bY_t=\bX_L\bH^T+\widetilde\bW,
\end{equation}
where $\widetilde\bW \in \bbC^{\tau_R \times
\Nt}$ is the AWGN matrix with elements being \iid $\mathcal{CN}(0,\sigma_{\tilde w}^2)$.

The reverse training signal sent by
LR allows the transmitter to obtain an estimate of the downlink channel by taking the transpose of its
uplink channel estimate. By employing the LMMSE estimator \cite{Est_Theory}, the estimate of $\bH$ at the transmitter can be written as
\begin{align}
\widehat{\mathbf{H}}=\left(\sH2\mathbf{X}_{L}^H(\sH2\mathbf{X}_{L}\mathbf{X}_{L}^H+
\sigma_{\tilde{w}}^2\mathbf{I}_{\tau_R})^{-1}\mathbf{Y}_{t}\right)^T \triangleq
\bH+\Delta\bH \label{equ.est.rev.channel}
\end{align}
where $\Delta\bH\in\bbC^{\Nt\times \NL}$ is the estimation
error matrix with
\begin{align}
\mathbb{E}\{\Delta\bH(\Delta\bH)^H\}\label{MSE.forward.channel.tx}
&=\NL\left(\frac{1}{\sigma_h^2}+\frac{\cE_R}{N_L\sigma_{\tilde{w}}^2}\right)^{-1}\bI_{\Nt},
\end{align}
and $\swt2$ is the noise power at the transmitter.

{\bf Step II (Forward Training with AN)}:
{After obtaining the downlink channel estimate, \ie $\widehat{\bH}$, in Step I, the transmitter
then sends a forward training signal to enable channel estimation at LR in Step II. To degrade the channel estimation performance of UR while not jamming LR, the transmitter
carefully inserts AN in the training signal.} The forward training signal is given by
\begin{equation}\label{eq.forward.training.AN}
{\bX_{t}=\sqrt{\frac{\cP_F\tau_F}{\Nt}}\bC_{t}+\bA\mathbf{K}_{\widehat{\bH}}^H},
\end{equation}
where $\bC_{t}\in \bbC^{\tau_F\times \Nt}$ is the pilot matrix with
$\text{Tr}(\bC_{t}^H\bC_{t})=\Nt$, $\cP_F$ is the pilot signal power in this stage, and $\tau_F$ is the
training length. For ease of notation, {we define $$\cE_F\triangleq \cP_F \tau_F$$ as} the forward
pilot signal energy. Here {$\bA \in \bbC^{\tau_F \times(\Nt-\NL)}$ is AN matrix whose entries are \iid
 $\mathcal{CN}(0,\sa2)$ and
are statistically independent of the channels and noises at all terminals;}
$\mathbf{K}_{\widehat{\bH}}\in\bbC^{\Nt\times(\Nt-\NL)}$ is a matrix whose column vectors form an
orthonormal basis for the left null space of $\widehat{\bH}$, that is,
$\mathbf{K}_{\widehat{\bH}}^H\widehat{\bH}=\mathbf{0}_{(\Nt-\NL)\times \NL}$ and
$\mathbf{K}_{\widehat{\bH}}^H\mathbf{K}_{\widehat{\bH}}=\bI_{\Nt-\NL}$. Notice from
(\ref{eq.forward.training.AN}) that AN is superimposed on the training signal and placed in the
left null space of $\widehat{\Hb}$ to minimize its interference on LR. The received signals at LR
and UR are respectively given by
\begin{align}\label{eq.receive.signal.LR}
\bY_{L}&=\sqrt{\frac{\cE_F}{\Nt}}\bC_{t}\bH+\bA\mathbf{K}_{\widehat{\bH}}^H\bH+\bW,\\
\bY_{U}&=\sqrt{\frac{\cE_F}{\Nt}}\bC_{t}\bG+\bA\mathbf{K}_{\widehat{\bH}}^H\bG+\bV, \label{eq.receive.signal.UR}
\end{align}
where $\bW\in\bbC^{\tau_F\times\NL}$ and $\bV\in\bbC^{\tau_F\times \NU}$ are the additive white Gaussian noise (AWGN)
matrices at LR and UR, respectively, with entries being \iid $\calC\calN(0,\sw2)$ and $\calC\calN(0,\sv2)$,
respectively.
Note that, since $\widehat\bH=\bH+\Delta\bH$ and $\mathbf{K}_{\widehat{\bH}}^H\mathbf{\widehat{H}}=\mathbf{0}$,
equation (\ref{eq.receive.signal.LR}) can be rewritten as
\begin{align} \label{equ.rec.LR}
\bY_L&=\sqrt{\frac{\cE_F}{\Nt}}\bC_t\bH-\bA\mathbf{K}_{\widehat{\bH}}^H\Delta\bH+\bW
\triangleq\bar{\mathbf{C}}\bH+\bar{\mathbf{W}},
\end{align}
where $\bar{\mathbf{C}}\triangleq \sqrt{\frac{\cE_F}{\Nt}}\bC_t$ and $\bar{\mathbf{W}}\triangleq
-\bA\mathbf{K}_{\widehat{\bH}}^H\Delta{\bH}+\bW$. {Also note that, due to the presence of $\Ab$, $\bar{\mathbf{W}}$ is statistically uncorrelated with $\bH$, i.e., $\E\{\bar{\mathbf{W}}^H\Hb\}=\zerob$.

By assuming the LMMSE estimator,} the channel estimate at LR can be expressed as
\begin{align}
\widehat{\mathbf{H}}_L=\bR_{\bH}{\bar\bC}^H(\bar\bC\bR_{\bH}{\bar\bC}^H+
\bR_{\bar \bW})^{-1}\mathbf{Y}_{L},\label{equ.est.for.channel}
\end{align}
where $\mathbf{R}_\bH=\mathbb{E}\{\bH\bH^H\}=\NL\sigma_h^2\bI_{\Nt}$ and $\mathbf{R}_{\bar{\bW}}=\mathbb{E}\{\bar\bW{\bar\bW}^H \}$.
The normalized mean squared error (NMSE) of $\mathbf{\widehat{H}}_L$
is defined as
\begin{align}\label{NMSE.forward.channel.LR}
\mbox{NMSE}_L&\triangleq
\frac{\mbox{Tr}\left(\mathbb{E}\{(\bH-\mathbf{\widehat{H}}_L)(\bH-\mathbf{\widehat{H}}_L)^H\}\right)}
{\Nt\NL}\notag\\
&=\frac{\mbox{Tr}\left(\left(\mathbf{R}_\bH^{-1}
+\bar{\mathbf{C}}^H\mathbf{R}_{\bar{\bW}}^{-1}\bar{\mathbf{C}}\right)^{-1}\right)}{\Nt\NL}.
\end{align}
By the fact that $\hat \bH$ and $\Delta\bH$ are uncorrelated due to the {orthogonality principle} \cite{Est_Theory}, and by (\ref{MSE.forward.channel.tx}) and
$\mathbf{K}_{\widehat{\bH}}^H\mathbf{K}_{\widehat{\bH}}=\bI_{\Nt-\NL}$, the covariance of $\bar\bW$ can be written as
\begin{align}
\mathbf{R}_{\bar{\bf W}}&=\left(\mathbb{E}\{\lVert\mathbf{K}_{\widehat{H}}^H\Delta\bH\rVert^2\}
\sigma_a^2+\NL\sigma_w^2 \right)\bI_{\tau_F}\notag\\
&=\NL\left[(\Nt-\NL)\cdot \left(\frac{1}{\sigma_H^2}+
\frac{\cE_R}{\NL\sigma_{\tilde{w}}^2} \right)^{-1}\sigma_a^2 +
\sigma_w^2\right]\bI_{\tau_F}.\label{var.noise.LR}
\end{align}
Then, by substituting (\ref{var.noise.LR}) into
(\ref{NMSE.forward.channel.LR}), we have
\begin{align}\label{NMSE2.forward.channel.LR}
&\mbox{NMSE}_L=\frac{1}{\Nt}\mbox{Tr}\bigg(\bigg(\frac{1}{\sigma_h^2}\bI_{\Nt}+
\notag \\
&~~~~~~~\frac{\cE_F}{\Nt}\frac{\bC^H_{t}\bC_t}
{(\Nt-\NL)\bigg(\frac{1}{\sigma_h^2}
+\frac{\cE_R}{\NL\sigma_{\tilde{w}}^2}\bigg)^{-1}\sigma_a^2+\sigma_w^2}\bigg)^{\!\!\!\!-1}\bigg).
\end{align}
Similarly,
the NMSE of the estimate of  $\bG$ at UR can be computed as
\begin{align}\label{NMSE2.forward.channel.UR}
&\mbox{NMSE}_U=\notag
\\
&~~\frac{1}{\Nt}\mbox{Tr}\left(\left(\frac{1}{\sigma_g^2}\bI_{\Nt}+\frac{\cE_F}{\Nt}\frac{\mathbf{C}^H_{t}\bC_t}
{(\Nt-\NL)\sigma_a^2\sigma_g^2+\sigma_v^2}\right)^{-1}\right).
\end{align}
Notice, from \eqref{NMSE2.forward.channel.LR} and \eqref{NMSE2.forward.channel.UR}, that increasing
the AN power (\ie $\sa2$) increases the NMSE at both receivers, but the effect can be reduced at LR
by increasing the reverse training energy $\cE_R$. Hence, under a total energy constraint, the
training and AN powers must be carefully chosen to ensure sufficient discrimination between the
channel estimation performances at the two receivers.

{\begin{remark}{\rm In view of the fact that the proposed forward training signal in
\eqref{eq.forward.training.AN} contains AN, an important question to ask is that whether UR can
ignore the AN-aided training signal and directly employ some blind data detection or channel
estimation methods in the data transmission phase. For example, if the transmitter uses space-time
coding schemes in the data transmission phase, UR may employ the blind detection methods in
\cite{Swindlehurst02,Ma06,Chang10} or the blind channel estimation methods in \cite{Shap05,Via08}.
However, these blind methods cannot work properly without cooperation of the transmitter.
Specifically, if the transmitter uses space-time codes that are not \emph{identifiable}
\cite{Via08,Ma07}, UR would suffer from nontrivial code and channel rotation ambiguities. It has
been shown in \cite{Hero03} that the rotation ambiguities can make UR have a detection error
probability equal to one, provided that the information symbols satisfy a constant modulus
property\footnote{In addition, the transmitter may employ the AN-aided OSTBC scheme in
\cite{Fakoorian2011} or the AN-aided beamforming schemes \cite{Swindlehurst,Swindlehurst2,Liao}.
Since the data signals of these schemes also contain AN, it would be even more difficult for UR to
use the blind methods to extract the information symbols or estimate the unknown channel.}. As a
result, UR may still need to exploit the training signal for channel estimation, even though it is
jammed by the AN signal. Further quantitative analysis evaluating the performance of other data transmission schemes and blind detection/channel estimation methods under
the proposed DCE scheme will be an interesting direction for future research.}
\end{remark}
}

\subsection{Optimal Power Allocation between Pilot Signal and AN}

The closed-form NMSE expressions obtained in the previous subsection show explicitly the effect of the
reverse training power (or energy), the forward training power (or energy) and the AN power on the
channel estimation performances at LR and UR. From a designer's point of view, it is
desirable to utilize the available power (or energy) in an efficient way whilst achieving a
satisfactory DCE performance. In the following, we consider the problem of allocating
the pilot signal and AN powers in the reverse and forward training stages with the goal of
minimizing the channel estimation error at the LR whilst keeping the estimation error at UR
above a certain threshold. The proposed optimization problem is given as follows:
{\begin{subequations}\label{opt.prob.I1}
\begin{align}
\min_{\cE_R,\cE_F,\sigma_a^2\geq0}&\mbox{NMSE}_L \\
\mbox{subject to (s.t.)}~~~ &\mbox{NMSE}_U \geq \gamma, \label{opt.prob.I1b}\\
&\cE_R \leq \bEL ,\label{opt.prob.I1d}\\
&\cE_F+(\Nt-\NL)\sigma_a^2\tau_F \leq \bEt.\label{opt.prob.I1e}
\end{align}
\end{subequations}
The optimization problem is constrained by a required lower limit on UR's NMSE in \eqref{opt.prob.I1b},
and individual energy constraints at LR and the transmitter in \eqref{opt.prob.I1d} and
\eqref{opt.prob.I1e}, respectively.}

It should be noted that UR's NMSE constraint, \ie $\gamma$, \cite{ChangChiangHongChi_TSP2010} should be chosen such that
\begin{equation}\label{cond.gamma}{
\left(\frac{1}{\sigma_g^2}+\frac{\bEt}
{\Nt\sigma_v^2}\right)^{-1} \leq \gamma \leq \sigma_g^2,}
\end{equation}
where the term on the left hand side is the minimum achievable NMSE at UR (when the transmitter
does not use any AN, \ie $\sigma_a^2=0$) and the term on the right hand side stands for the worst
NMSE performance at UR (which is achieved when the mean of $\Gb$, i.e., zero, is taken as the channel estimate). For ease of use later, let us
define
\begin{equation} \tgam \triangleq \left(
\frac{1}{\gamma}-\frac{1}{\sigma_g^2}\right)\Nt\sigma_v^2\geq 0
\end{equation}
so that the condition in \eqref{cond.gamma} can be reduced to
\begin{equation} \label{cond.tigamma}
0 \leq \tgam \leq \bEt.
\end{equation}

The power allocation problem in \eqref{opt.prob.I1} is a non-convex optimization problem involving
three variables ($\cE_R,\cE_F,\sigma_a^2$). Interestingly, we show in the following proposition that, for the case of orthogonal forward pilot
matrices (\ie the case where $\bC_t^H\bC_t=\bI_{\Nt}$), {problem \eqref{opt.prob.I1}
actually has simple closed-form solutions.}

{
\begin{Prop}\label{prop.optpower.rec} Consider problem \eqref{opt.prob.I1} with $\gamma$ chosen according to \eqref{cond.gamma}.
If 
\begin{align} \label{condition u}
\mu \triangleq \NL
\left(\frac{\sigma_v^2\sigma_{\tilde{w}}^2}{\sigma_g^2\sigma_w^2}
-\frac{\sigma_{\tilde{w}}^2}{\sigma_h^2}\right)
>\bEL,
\end{align}
the optimal $(\cE_R,\cE_F,\sa2)$ is given by $\cE_R^\star=0$,
$\cE_F^\star=\tgam$ and $(\sa2)^\star=0$ (\ie no need of reverse training and AN). On the other hand, if
$\mu\leq\bEL$, the transmitter and LR use all the available energy so that
$\cE^\star_R=\bEL$,
$$\cE^\star_F=\bEt-\frac{(\bEt-\tgam)\tau_F}{\tau_F+\tgam\sG2/\sv2},$$
and
$$(\sa2)^\star=\frac{\bEt-\tgam}{\left(\tau_F+\tgam\sG2/\sv2\right)(\Nt-\NL)}.$$
\end{Prop}

Proposition~\ref{prop.optpower.rec} implies that, if UR has a relatively poor channel condition
(i.e., sufficiently small ${\sigma_g^2}/{\sigma_v^2}$), then both AN and reverse training are
not needed; otherwise, LR should use all its energy for reverse training and the transmitter needs
to employ AN in order to constrain UR's MSE above the required threshold value $\gamma$. The
proof of Proposition \ref{prop.optpower.rec} is given in Appendix~\ref{append.prop1}.
Appendix~\ref{append.prop1} actually provides the proof for a more general formulation which,
compared to problem \eqref{opt.prob.I1}, has an additional total energy constraint
$$
\cE_R +\cE_F+(\Nt-\NL)\sigma_a^2\tau_F \leq \bEave,
$$
where $\bEave$ represents the total energy budget. This total energy constraint limits the total amount of energy consumed by the transmitter and LR in the training phase. We are interested in such general formulation because it
 may be useful in the system design stage for understanding the power tradeoff between the transmitter and LR and that between the training phase and data transmission phase.
}



Two remarks regarding the proposed DCE scheme are in order.

\begin{remark}\label{rmk.training_design}
{\rm It is interesting to note that Proposition~\ref{prop.optpower.rec} gives the solution to the
optimization problem for the orthogonal forward training matrix with full rank, \ie
$\bC^H_t\bC_t=\bI_{\Nt}$. However, the rank of $\bC_t$ does not need to be $\Nt$ in general. Given
that the rank of $\bC_t$ is equal to $K (<\Nt)$, it is shown in Appendix~\ref{append.optbCt} that
the optimal $\bC_t$ must satisfy $\bC^H_t\bC_t=\mathbf{U}_c\mathbf{DU}_c^H$, where
$\mathbf{D}=\diag(d_1,\dots,d_K,$ $0,\dots,0)$ with $d_1=\dots=d_K=\frac{\Nt}{K}$ and
$\mathbf{U}_c$ is an $\Nt\times \Nt$ unitary matrix.
If a rank
deficient pilot matrix is considered instead, \ie $K < \Nt$, one can choose an arbitrary
$\Nt\times\Nt$ unitary matrix for $\mathbf{U}_c$ and obtain the optimal $(\cE_R,\ \cE_F,\ \sa2)$
for a given $K$ value by following the same derivations as in the proof of Proposition \ref{prop.optpower.rec}. The rank of the forward training matrix can be further optimized to minimize the NMSE at LR.}
\end{remark}

\begin{remark}\label{rmk.training_length}
{\rm The training lengths in the reverse and forward training stages, \ie $\tau_R$ and $\tau_F$,
can also be optimized. Note that training on the reverse link is not affected by the presence of UR
and, thus, can be viewed as training on a point-to-point link. Therefore, by
\cite{Hassibi_howmuchtraining}, the optimal reverse training length $\tau_R$ is equal to the number
of antennas at LR (\ie $\NL$) since it minimizes the training overhead without compromising the
channel estimation performance. {One can also show that the optimal forward training length
$\tau_F$ is given by the number of antennas of the transmitter, i.e., $\Nt$. To show this, observe
that the optimal $(\cE_R^\star,\ \cE_F^\star,\ \sa2^\star)$ of (\ref{opt.prob.I1}) for some
$\tau_F=\tau_{F_2}$ satisfies the constraints of (\ref{opt.prob.I1}) even when $\tau_F$ reduces to
a smaller value $\tau_{F_1}<\tau_{F_2}$, i.e., $(\cE_R^\star,\ \cE_F^\star,\ \sa2^\star)$ is also
feasible to (\ref{opt.prob.I1}) with $\tau_F=\tau_{F_1}$. This implies that a smaller $\tau_F$
corresponds to a larger feasible set for $(\cE_R,\ \cE_F,\ \sa2)$ in (\ref{opt.prob.I1}), and thus
a smaller value of optimal ${\rm NMSE}_L$ can be obtained. Consequently, the minimum ${\rm NMSE}_L$
is achieved when $\tau_F$ is equal to its minimum possible value, \ie the number of antennas at the
transmitter $\Nt$.} }
\end{remark}

\section{Two-Way DCE Design for Non-Reciprocal Channels}\label{ch4.NonRecip}

In this section, we consider the case of non-reciprocal channels. Without channel reciprocity, the
downlink channel gain cannot be directly inferred from the uplink channel gain. In this case, the
knowledge of the downlink channel at the transmitter can be obtained via reverse training plus an
additional round-trip training stage that utilizes an echoed signal (from the transmitter to LR and
back to the transmitter) to obtain the combined downlink-uplink channel at the transmitter. The
proposed two-way DCE training scheme is detailed below.

\subsection{Training Scheme and Channel Estimation Performance}\label{ch4.NonRecip-A}

{\bf Step I (Channel Acquisition at the Transmitter with Round-Trip and Reverse Training)}: In the
round-trip training stage, the transmitter first broadcasts a random signal, only known to itself,
which is then echoed back to the transmitter by LR. The effective channel seen at the transmitter
is equal to the composition of the uplink and downlink channels. Specifically, let $\bC_{t0} \in
\bbC^{\tau_{t0} \times \Nt}$ be the pilot matrix that is randomly generated with normalized power,
\ie $\text{Tr}(\bC_{t0}^H\bC_{t0})=\Nt$. The signal sent by the transmitter is given by
\begin{equation}
 \bX_{t0}=\sqrt{\frac{{\cP_{t0} \tau_{t0}}}{\Nt}}\bC_{t0},
\end{equation}
where $\cP_{t0}$ is the pilot signal power and $\tau_{t0}$ is the training length in this stage.
{Note that $\bC_{t0}$ is known only to the transmitter (but not to LR or UR\footnote{UR may attempt
to exploit the AN-free signal $\bC_{t0}$ to obtain some useful information about its channel $\Gb$.
For example, if $N_U>\Nt$ or the distribution of $\bC_{t0}$ is known, UR can estimate the subspace
of $\Gb$ from the received signal. However, like the blind methods discussed Remark 1, this
subspace information still suffers from a nontrivial rotation ambiguity about the true channel $\Gb$.
Besides, the subspace estimation quality could be very poor due to the short length of $\bC_{t0}$
and the presence of additive noise. }) since it is randomly generated before each transmission.}
For ease of notation, we define the round-trip training energy as {$\cE_{t0}\triangleq \cP_{t0}
\tau_{t0}$.} The received signal at LR is given by
\begin{equation}
\bY_{L0}=\bX_{t0}\bHd+\bW_0,
\end{equation}
where $\bW_0\in \bbC^{\tau_{t0}\times \NL}$ is the AWGN matrix with entries that are \iid with distribution $\calC\calN(0,\sw2)$. Upon receiving $\bY_{L0}$, LR amplifies and
forwards it back to the transmitter. The echoed signal at the transmitter is given by
\begin{align}\label{eq.echoed.signal}
\bY_{t1}&=\alpha\bY_{L0}\bHu+\bWt_1 \notag\\
     &=\alpha\bX_{t0}\bHd\bHu+\alpha\bW_0\bHu+\bWt_1,
\end{align}
where $\bWt_1\in \bbC^{\tau_{t0} \times \Nt}$ is the AWGN matrix at the transmitter with entries being \iid with distribution $\calC\calN(0,\swt2)$. The amplifying gain at LR is given by
\begin{align}
\alpha&=\sqrt{\frac{{\cP_{L1}\tau_{t0}}}{\cP_{t0} { \tau_{t0}}\NL\sHd2+\tau_{t0} \NL \sw2}}\notag \\
&=\sqrt{\frac{{ \cE_{L1}}}{\cE_{t0}\NL\sHd2+{\tau_{t0}}\NL\sw2}},
\end{align}
where {$\cP_{L1}$} is LR's transmission power and {$\cE_{L1}\triangleq \cP_{L1} \tau_{t0}$} is the
energy spent on echoing the signal. With the knowledge of $\bX_{t0}$, the transmitter can obtain an
estimate of the combined downlink and uplink channels, \ie $\bHd\bHu$. However, to obtain an
estimate of the downlink channel $\bHd$, the transmitter must first obtain an estimate of the
uplink channel $\bHu$. This can be achieved through reverse training, which is the same as the one
described in the reciprocal channel case.

In the reverse training stage, LR sends a training signal
$\bX_{L2}=\sqrt{\frac{\cE_{L2}}{\NL}}\bC_{L2} \in \bbC^{\tau_{L2}\times \NL}$ to enable uplink
channel estimation at the transmitter. Here, $\bC_{L2}$ is the pilot matrix satisfying
$\bC^H_{L2}\bC_{L2}=\bI_{\NL}$, {$\cE_{L2}$} is the reverse training energy, and {$\tau_{L2}$} is
the training length. The received signal at the transmitter is given by
\begin{equation}
\bY_{t2}=\bX_{L2}\bHu+\bWt_2=\sqrt{\frac{\cE_{L2}}{\NL}}\bC_{L2}\bHu+\bWt_2,
\end{equation} where $\bWt_2$ is the AWGN matrix with entries being \iid with distribution $\calC\calN(0,\swt2)$.
The transmitter can then obtain the LMMSE estimate of the uplink channel as
\begin{align}
\widehat{\bH}_u=\sHu2\mathbf{X}_{L2}^H(\sHu2\mathbf{X}_{L2}\mathbf{X}_{L2}^H+
\sigma_{\tilde{w}}^2\mathbf{I}_{\tau_{L2}})^{-1}\mathbf{Y}_{t2}. \label{equ.est.rev.channel.nonrec}
\end{align}
Similar to that in \eqref{equ.est.rev.channel} and \eqref{MSE.forward.channel.tx}, we can write
\begin{equation}\label{eq.HuEst.nonreciprocal}
\widehat{\bH}_u\triangleq \bH_u+\Delta\bH_u,
\end{equation}
where {$\Delta\bH_u$ is the
estimation error matrix which is complex Gaussian distributed with zero mean and correlation matrix}
\begin{equation}\label{eq.mse.uplinkbHu}
\mathbb{E}\{\Delta\bH_u^H(\Delta\bH_u)\}
=\NL\left(\frac{1}{\sHu2}+\frac{\cE_{L2}}{N_L\sigma_{\tilde{w}}^2}\right)^{-1}\bI_{\Nt}.
\end{equation}
{Note that $\Delta\bH_u$ and $\widehat{\bH}_u$ are statistically independent since they are both Gaussian and are uncorrelated with each other due to the orthogonality principle \cite{Est_Theory}.}
The transmitter can then utilize the uplink channel estimate $\widehat\bH_u$ to compute an estimate of the downlink channel $\bHd$.

Specifically, given $\hbHu$ at the transmitter, we can
rewrite the echoed signal in (\ref{eq.echoed.signal}) as
\begin{align}
\bY_{t1}&=\alpha \bX_{t0}\bHd(\hbHu-\Delta\bHu)+\alpha\bW_0(\hbHu-\Delta\bHu)+\bWt_1 \notag \\
&=\alpha \bX_{t0}\bHd\hbHu+(\alpha\bW_0\hbHu-\alpha\bX_{t0}\bHd\Delta\bHu \notag \\
&~~~~-\alpha\bW_0\Delta\bHu+\bWt_1).\label{eq.echoed.signal.givenhbHu}
\end{align}
For ease of analysis, let us define $\mathbf{y}_{t1}=\text{vec}(\bY_{t1})$, $\mathbf{h}_{d}=\text{vec}(\bHd)$, $\mathbf{w}_{0}=\text{vec}(\bW_0)$, and $\mathbf{\tilde{w}}_1=\text{vec}(\bWt_1)$ as the respective vector counterparts of $\bY_{t1}$, $\bHd$,
$\bW_0$ and $\bWt_1$ obtained by stacking the column vectors of each corresponding matrix. By the Kronecker product property that $\text{vec}(\mathbf{ABC})=(\mathbf{C}^T\otimes \mathbf{A})\text{vec}(\mathbf{B})$, one can express (\ref{eq.echoed.signal.givenhbHu}) as
\begin{align}\label{temp}
\mathbf{y}_{t1}&=\alpha(\hbHu^T\otimes \bX_{t0})\mathbf{h}_d
+\alpha(\hbHu^T\otimes \bI_{\Nt})\mathbf{w}_0\notag \\
&-\alpha(\Delta\mathbf{H}_u^T\otimes \bX_{t0})\mathbf{h}_d-\alpha(\Delta\mathbf{H}_u^T\otimes \bI_{\Nt})\mathbf{w}_0
+\mathbf{\tilde{w}}_1.
\end{align}

Let $\bC_{t0}$ be a unitary matrix such that  $\bC_{t0}\bC^H_{t0}=\bC^H_{t0}\bC_{t0}=\bI_{\Nt}$\footnote{Note that the pilot matrix $\bC_{t0}$ need not be a unitary matrix in general. However, the NMSE performance obtained with a generic round-trip pilot
matrix cannot be expressed in a closed form and, thus, the optimal pilot structure is difficult to
obtain. In this work, we aim to provide a design that can be efficiently implemented and whose
performance can be easily characterized. With the choice of unitary pilot matrices, we are able to derive an accurate closed-form approximation of the
NMSE performance and further utilize it to efficiently optimize the power
(energy) allocation among the pilot signals and AN in different stages.}.
{By the fact that the equivalent noise term $\alpha(\hbHu^T\otimes \bI_{\Nt})\mathbf{w}_0-\alpha(\Delta\mathbf{H}_u^T\otimes \bX_{t0})\mathbf{h}_d-\alpha(\Delta\mathbf{H}_u^T\otimes \bI_{\Nt})\mathbf{w}_0
+\mathbf{\tilde{w}}_1$ in \eqref{temp} is statistically uncorrelated with $\mathbf{h}_d$, the LMMSE estimate}
of the downlink channel $\mathbf{h}_d$ at the transmitter (denoted by $\mathbf{\hat{h}}_{d,t}$) can be computed as
\begin{align}
\mathbf{\hat{h}}_{d,t}
&=\frac{1}{\alpha\sw2}\bigg(\frac{1}{\sHd2}+\frac{\cE_{t0}}{\Nt\sw2}\bigg)^{-1}\notag \\
&~~~~\times \bigg(\hbHu^*
\bigg( (\hbHu^T\hbHu^*)+\beta\bI_{\Nt}\bigg)^{-1}
\otimes \bX_{t0}^H\bigg)\mathbf{y}_{t1} \notag \\
&\triangleq\mathbf{h}_d+\Delta\mathbf{h}_{d,t}, 
\end{align}
where \begin{equation}\label{eq.beta}
\beta= \NL\left( \frac{1}{\sHu2}+\frac{\cE_{L2}}{\NL\swt2}\right)^{-1}\!\!\!+
\frac{\swt2}{\alpha^2\sHd2\sw2}\left(\frac{1}{\sHd2}+\frac{ \cE_{t0}}{\Nt\sw2}\right)^{-1}.
\end{equation}
and $\Delta\mathbf{h}_{d,t}\in\bbC^{\Nt\NL\times 1}$ is the estimation error
vector at the transmitter.
The corresponding matrix form of $\mathbf{\hat{h}}_{d,t}$ is given by
\begin{align}\label{hatHdt}
\hbHdt&=\frac{1}{\alpha\sw2}\left(\frac{1}{\sHd2}+\frac{ \cE_{t0}}{\Nt\sw2}\right)^{-1}
\notag \\
&~~~~\times\bX_{t0}^H\bY_{t1}\left( (\hbHu^H\hbHu)+\beta\bI_{\Nt}\right)^{-1}\hbHu^H \notag \\
&\triangleq\bHd+\Delta\mathbf{H}_{d,t},
\end{align}
where $\bHd$ and $\Delta\bH_{d,t}$ is the matrix form of $\mathbf{h}_d$ and
$\Delta\mathbf{h}_{d,t}$, respectively. The covariance matrix of $\Delta\mathbf{h}_{d,t}$
conditioned on $\hbHu$ is given by
\begin{align}\label{eq.mse.downlinkHd.tx}
&\mathbb{E}\{\Delta\mathbf{h}_{d,t}(\Delta\mathbf{h}_{d,t})^H|\hbHu\}
=\bigg[\sHd2\bI_{\NL}-\sHd2\frac{\sHd2 \cE_{t0}}{\sHd2 \cE_{t0}+\Nt\sw2} \notag \\
&~~~~\times\bigg(\!
\bigg(\frac{1}{\beta}\hbHu^*\hbHu^T\bigg)^{-1}\!\!\!\!+\!\bI_{\NL}\!\bigg)^{-1}\bigg]\otimes \bI_{\Nt}.
\end{align}

{\bf Step II (Forward Training with AN)}: In the forward training stage, the transmitter sends AN along with the training signal to discriminate the channel estimation performances between LR
and UR. The detailed description has been presented earlier in Section II-B. For notational
consistency in this section, we modify the subscripts of the symbols in the expressions of the
received signals at LR and UR as
\begin{align}
\bY_{L3}&=\sqrt{\frac{{ \cE_{t3}}}{\Nt}}\bC_{t3}\bHd+\bA\bNHdHt\bHd+\bW_3,\label{eq.receive.signal.LR3}\\
\bY_{U3}&=\sqrt{\frac{\cE_{t3}}{\Nt}}\bC_{t3}\bG+\bA\bNHdHt\bG+\bV_3, \label{eq.receive.signal.UR3}
\end{align}
and the forward training length is denoted as {$\tau_{t3}$} instead of $\tau_F$.

Comparing with the design for reciprocal channels described in Section~\ref{ch3.Recip}, the DCE in
the case of non-reciprocal channels requires more time to complete since an additional round-trip
training stage is used. To keep the training overhead low, we consider a design with the minimum
training lengths, \ie $\tau_{t0} = \Nt$, $\tau_{L2}=\NL$, $\tau_{t3}=\Nt$, and choose $\bC_{t0}$ to
be a unitary matrix such that $\bC^H_{t0}\bC_{t0}=\bC_{t0}\bC^H_{t0}=\bI_{\Nt}$. Note that the
choices of $\tau_{L2}=\NL$ and $\tau_{t3}=\Nt$ are indeed optimal under a total and/or individual
energy constraints as discussed in Remark~\ref{rmk.training_length} of Section~\ref{ch3.Recip}.



{Unlike the reciprocal channel case in Section \ref{chap2-2}, it is difficult to obtain a
close-form expression for the NMSE at LR from \eqref{eq.receive.signal.LR3}. In Appendix
\ref{append.deriv.nonNMSEL}, we instead derive an approximate NMSE at LR by assuming that: 1)
given $\widehat{\Hb}_u$, the LMMSE estimate of $\Hb_d$ in \eqref{hatHdt}, i.e., $\widehat{\Hb}_{d,t}$, is statistically independent of the associated error matrix
$\Delta{\Hb}_{d,t}$; 2) the transmitter and LR have sufficiently large numbers of antennas, i.e,
$\Nt,N_U \gg 1$. The obtained approximation of the NMSE at LR is given by \eqref{eq.NMSEL.appr2} which is at the top of the next page,}
\begin{figure*}[t!]
\begin{align}\label{eq.NMSEL.appr2}
\text{NMSE}_L\approx
\frac{1}{\Nt}\mbox{Tr}\left(\frac{1}{\sHd2}\bI_{\Nt}+\frac{\cE_{t3}}{\Nt}\frac{\bC^H_{t3}\bC_{t3}}
{(\Nt-\NL)\sa2\left(\sHd2-\sHd2\frac{\sHd2 \cE_{t0}}{\sHd2 \cE_{t0}+\Nt\sw2}
\frac{\Nt\sigma^2}{\beta+\Nt\sigma^2}\right)+\sw2} \right)^{-1}
\end{align}
\hrulefill\vspace{.0cm}
\end{figure*}
where
\begin{align}\label{eq.sigma}
\sigma^2\triangleq \frac{\sigma_{h_u}^4\cE_{L2}}{\sHu2\cE_{L2}+\NL\swt2},
\end{align}
and $\beta$, as recalled from \eqref{eq.beta}, is a function of the energy values $\cE_{t0}$ and
$\cE_{L2}$. {While the two assumptions are in general not true (in general $\widehat{\Hb}_{d,t}$
and $\Delta{\Hb}_{d,t}$ are only statistically uncorrelated, and $\Nt$ and $N_U$ are finite), our
numerical results show that the approximate $\text{NMSE}_L$ presented above is actually quite
accurate under practical settings, as will be shown in the numerical result section.

The NMSE performance at UR can be computed in a similar fashion as in the reciprocal case, which is
given by}
\begin{align}
&\text{NMSE}_U
=\notag\\
&~~\frac{1}{\Nt}\mbox{Tr}\left(\left( \frac{1}{\sG2}\bI_{\Nt}+\frac{\cE_{t3}}{\Nt}\frac{\bC^H_{t3}\bC_{t3}}
{(\Nt-\NL)\sa2\sG2+\sv2}\right)^{-1}\right).\label{eq.NMSEU}
\end{align}

\subsection{Optimal Power Allocation between Training and AN Signals}

The effect of power allocation on the NMSE performance is much more complex in the non-reciprocal
case.
Nonetheless, we can formulate an optimization problem, similar to that in the reciprocal case, where we aim to
 minimize the channel estimation error at LR whilst
keeping the estimation error at UR above a threshold.
The optimization problem is given as follows:{
\begin{subequations}\label{prob.Nonrecip}
\begin{align}
\underset{\substack{\cE_{t0},\cE_{L1},\\\cE_{L2},\cE_{t3},\sa2\geq 0}}{\min}\ \ &\text{NMSE}_L\\
\text{s.t.}\ \ \ \ \ \ \ \ &\text{NMSE}_U\geq \gamma\\
&\cE_{t0}+\cE_{t3}+(\Nt-\NL)\sa2 \Nt\leq \bEt\\
&\cE_{L1}+\cE_{L2}\leq \bEL.
\end{align}
\end{subequations}
Here, $\bEt$ and $\bEL$ are the individual energy constraints at
the transmitter and LR, respectively.} Since this problem is not easily solvable, we resort to
the monomial approximation and the condensation method often adopted in the field of
GP{\cite{Tutorial_GP2,Tutorial_GP}} to obtain an efficient solution. Detailed description of the numerical
algorithm can be found in~\cite{ref.Condensation} and are omitted in this paper since these approaches are standard in the field of GP.

\begin{remark} {\rm Similar to the discussion in Remark \ref{rmk.training_design} of Section~\ref{ch3.Recip},
one can also show that the optimal structure of $\bC_{t3}$ is given by
$\bC^H_{t3}\bC_{t3}=\mathbf{U}_{t3}\mathbf{D}\mathbf{U}^H_{t3}$, where
$\mathbf{D}=\diag(d_1,\dots,d_K,$ $0,\dots,0)$ with $d_1=\dots=d_K=\frac{\Nt}{K}$ and
$\mathbf{U}_{t3}$ is the matrix whose columns consist of eigenvectors of $\bC^H_{t3}\bC_{t3}$. For
any choice of $K (\leq\Nt)$, the optimal power allocation can be obtained by performing the same monomial
approximation and condensation method described in~\cite{ref.Condensation}. The optimal $K$ can then be found by comparing the solutions for all possible values of $K$.
}
\end{remark}

{
\begin{remark} {\rm As the final remark, it would be interesting to qualitatively compare the proposed two-way DCE scheme and the original feedback-and-retraining DCE
scheme in \cite{ChangChiangHongChi_TSP2010}. In terms of training overhead, it is easy to see that
the proposed two-way DCE scheme is more efficient than the feedback-and-retraining DCE scheme,
since the two-way DCE scheme requires  at most three transmissions by the transmitter and/or LR
(e.g., for the non-reciprocal channels, we require one round-trip transmission, one reverse
training and one forward training with AN) while the feedback-and-retraining DCE scheme, even under
the assumption of ideal feedback, usually requires around five stages of feedback and retraining
(equivalent to 10 transmissions by the transmitter and LR) in order to achieve a comparable
performance \cite{ChangChiangHongChi_TSP2010}. We should mention that, if the goal is solely to
prevent UR from obtaining a good estimate of its channel from the transmitter, the
feedback-and-retraining scheme actually provides no advantages over the two-way DCE scheme, even
though it requires more complex operations. However, the two-way DCE scheme cannot be applied if
the channel between LR and UR is also of interest at UR, e.g., when LR also has a secret message to
transmit. In this scenario, the feedback-and-retraining scheme in \cite{ChangChiangHongChi_TSP2010}
would be the preferred method. In summary, the two DCE schemes actually have their own values and
limitations and should be deployed depending on the applications. }
\end{remark}
}

\section{Numerical Results and Discussions}\label{ch7.simulation}


In this section, we present numerical results to demonstrate the effectiveness of the proposed DCE
schemes. We consider a MIMO wireless system as described in Section \ref{ch2.system} with $\Nt=4$,
$\NL=2$ and $\NU=2$. The elements of the channel matrices, $\bH$, $\bHu$, $\bHd$ and $\bG$, are
\iid complex Gaussian distributed with zero mean and unit variance, \ie $\sH2=\sHu2=\sHd2=\sG2=1$.
The entries of the receiver noise matrices, \ie $\bWt$, $\bW$ and $\bV$, are also assumed to be
\iid complex Gaussian with zero mean and unit variance, \ie
$\sigma_{\tilde{w}}^2=\sigma_w^2=\sigma_v^2=1$. The orthogonal forward pilot matrix is employed,
\ie $\bC^H_t\bC_t=\bC^H_{t3}\bC_{t3}=\bI_{\Nt}$. Moreover, the training lengths are set to its
minimum, that is, $\tau_R=N_L=2$ and $\tau_F=\Nt=4$ for the reciprocal case, and
$\tau_{t0}=\tau_{t3}=\Nt=4$ and $\tau_{L2}=N_L=2$ for the non-reciprocal case. Let us denote the
total training length spent by the transmitter as $\tau_t$ and that by LR as $\tau_L$. For the
reciprocal case, $\tau_t=4~(=\tau_F)$, $\tau_L=2~(=\tau_R)$, and for the non-reciprocal case,
$\tau_t=8~(=\tau_{t0}+\tau_{t3})$, $\tau_L=6~(=\tau_{t0}+\tau_{L2})$. Note that the overall
training time would be longer than the sum of all training lengths due to the processing time at
the transmitter and LR. {In order to investigate the energy tradeoff between the transmitter and LR, we consider the general formulation as in \eqref{opt.prob.I12} which has the additional total energy constraint in contrast to \eqref{opt.prob.I1} and \eqref{prob.Nonrecip}.}
We define the average transmit power as
$\bPave\triangleq\frac{\bEave}{\tau_t+\tau_L}$ so that a total energy budget can be alternatively
expressed in terms of an average power budget. For all simulation results, the individual power
constraints at the transmitter and LR are respectively given by
$\bPt\triangleq\frac{\bEt}{\tau_t}=30$~dB and $\bPL\triangleq\frac{\bEL}{\tau_L}=20$~dB (relative
to its noise variance). We incorporate an NMSE lower bound for comparison. The lower bounds for the
reciprocal and the non-reciprocal cases are both given by
\begin{equation}
\mbox{NMSE}_{\mbox{LB}}=\left( \frac{1}{\sigma_{H_{(d)}}^2}+\frac{\min\{\bEt ,
\ \bEave\}}{\Nt\sigma_w^2}\right)^{-1}.
\end{equation}
This is the minimum achievable NMSE at LR when $\sigma_a^2=0$, \ie no AN is used.

\begin{figure}[t!]
\begin{center}
\subfigure[Reciprocal case]{
\psfrag{Pave}[Bc][Bc]{ $\bPave$ (dB)}
\psfrag{forwardgamma01forward}[Bl][Bl]{{\small $\mathcal{E}_F/\tau_F$,~$\gamma=0.1$}}
\psfrag{reversegamma01}[Bl][Bl]{\small $\mathcal{E}_R/\tau_R$,~$\gamma=0.1$}
\psfrag{angamma01}[Bl][Bl]{\small $(\Nt-N_L)\sigma_a^2$,~$\gamma=0.1$}
\psfrag{forwardgamma003}[Bl][Bl]{{\small $\mathcal{E}_F/\tau_F$,~$\gamma=0.03$}}
\psfrag{reversegamma003}[Bl][Bl]{\small $\mathcal{E}_R/\tau_R$,~$\gamma=0.03$}
\psfrag{angamma003}[Bl][Bl]{\small $(\Nt-N_L)\sigma_a^2$,~$\gamma=0.03$}
\psfrag{Powervalue}[Bc][Bc]{dB}

\label{fig-PowAllo_r01003}
\includegraphics[scale=0.57]{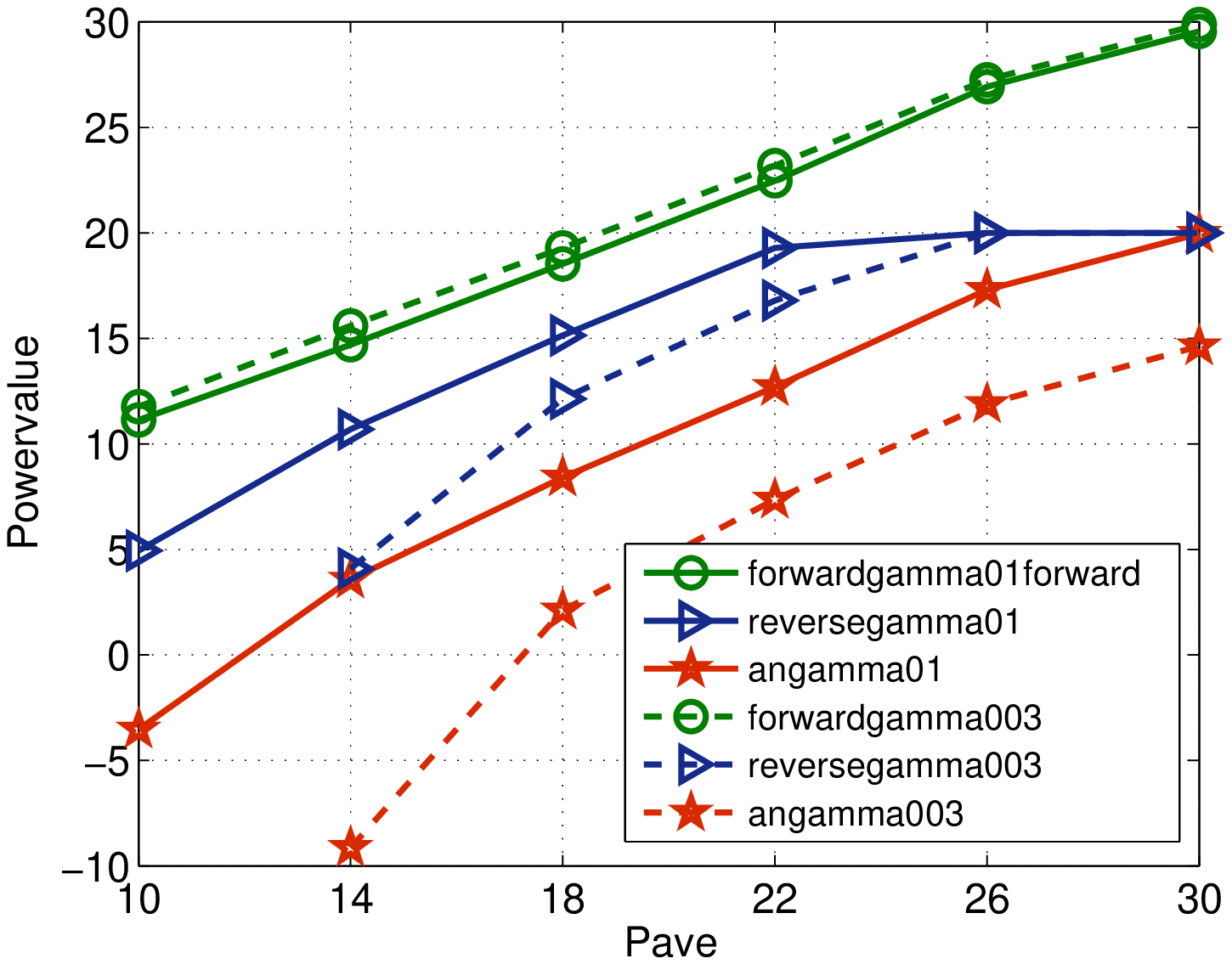}}

\subfigure[Non-reciprocal case]{
\psfrag{Pave}[Bc][Bc]{ $\bPave$ (dB)}
\psfrag{powervalue}[Bc][Bc]{dB}
\psfrag{aaaaaaaaaaa}[Bl][Bl]{{\small $\mathcal{E}_0/\tau_{t0}$}}
\psfrag{bb}[Bl][Bl]{{\small $\mathcal{E}_1/\tau_1$}}
\psfrag{cc}[Bl][Bl]{{\small $\mathcal{E}_2/\tau_{L2}$}}
\psfrag{dd}[Bl][Bl]{{\small $\mathcal{E}_3/\tau_{t3}$}}
\psfrag{ee}[Bl][Bl]{{\small $(\Nt-N_L)\sigma_a^2$}}
\psfrag{ff}[Bl][Bl]{{\small $\gamma=0.03$}}
\psfrag{gg}[Bl][Bl]{{\small $\gamma=0.1$}}

\label{fig-PowAllononrecip}
\includegraphics[scale=0.57]{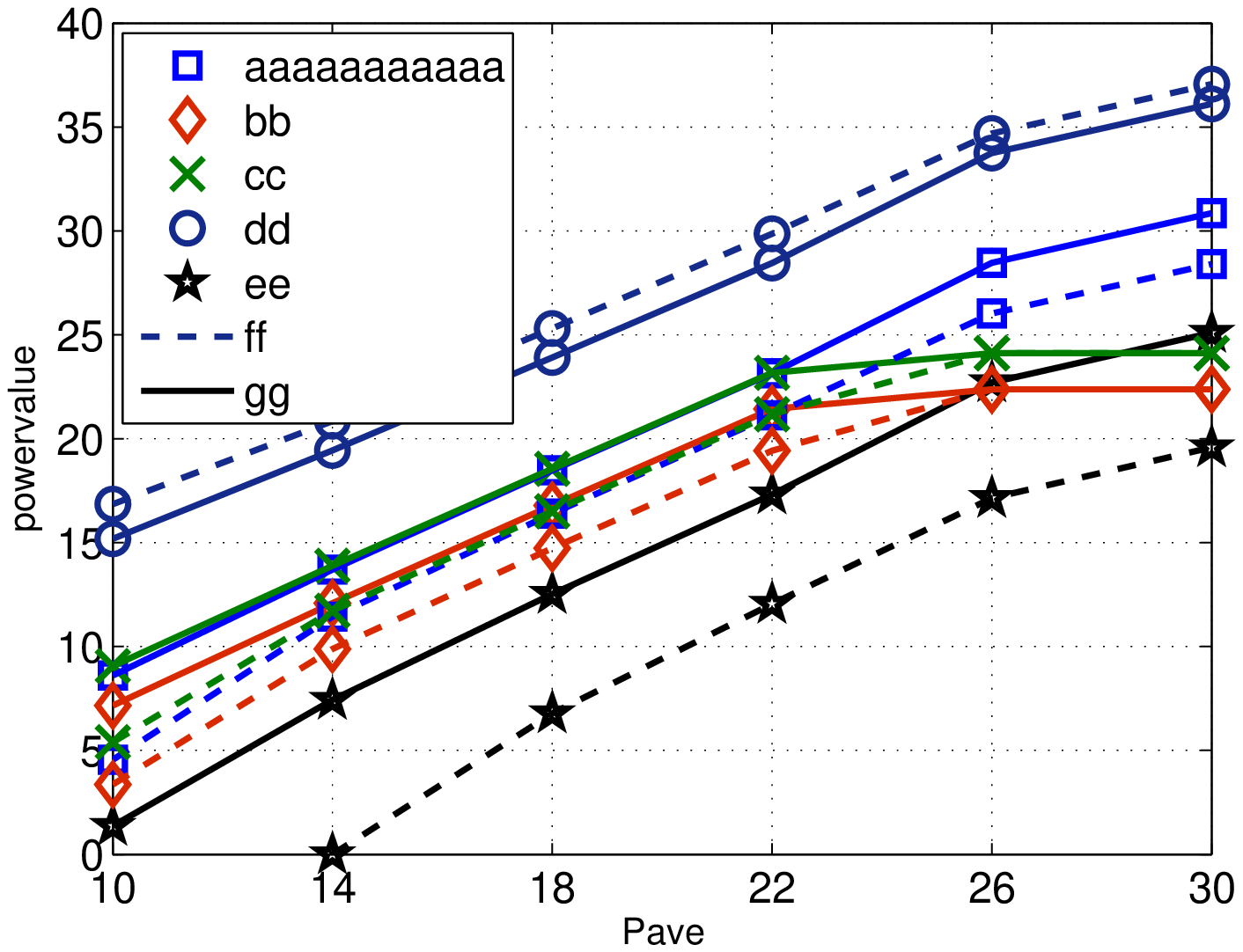}}
\caption{Power allocation between the pilot signal and AN powers of the proposed DCE schemes for the reciprocal and the non-reciprocal cases.}
\label{fig-NMSE1}
\end{center}
\vspace{-0.5cm}
\end{figure}


In Fig.~\ref{fig-NMSE1}, we show the results of the optimal power allocation among pilot and AN
signals in different stages of the training process for both the reciprocal and the non-reciprocal
cases. The results are shown for two different lower limits on UR's NMSE, \ie $\gamma=0.1$
(indicated by solid lines) and $\gamma=0.03$ (indicated by dashed lines). In the reciprocal case,
the reverse training power, forward training power, and the AN power are defined as $\cE_R/\tau_R$,
$\cE_F/\tau_F$ and $(\Nt-\NL)\sigma_a^2$, respectively. We can see from
Fig.~\ref{fig-PowAllo_r01003} that, as the average power budget $\bPave$ increases, all the
training powers and the AN power increase at roughly the same rate over the range of $\bPave$ from
18~dB to 26~dB. This suggests that the optimal percentages of total energy allocated to the reverse
training, forward training, and AN do not change much over a wide range of energy budget. However,
when $\bPave$ becomes sufficiently high, the curves in Fig.~\ref{fig-PowAllo_r01003} start
flattening out due to the individual power constraints at the transmitter and LR.
Moreover, by comparing the optimal power allocation of the case with $\gamma=0.03$ and that with
$\gamma=0.1$, we can see that it is desirable to allocate more power to AN and less power to the
forward pilot signal as $\gamma$ increases (\ie when a stricter constraint is imposed on UR's
performance). This is due to the fact that the forward pilot signal benefits both LR and UR while
AN primarily degrades UR's estimation. However, an increase in the AN power may also degrade
the estimation performance at LR if it is not placed accurately in the null space of the channel to
LR. To reduce this effect, the reverse training power should be increased in order to obtain a more
accurate knowledge of the downlink channel. This explains why the reverse training power increases
with $\gamma$ in Fig. \ref{fig-PowAllo_r01003}. Note that, when $\bPave$ falls below $10$ dB, the
value of $\gamma=0.03$ falls out of the feasible range given in (\ref{cond.gamma}) (because the
value of $\gamma$ is not achievable by UR even without AN) and, therefore, is not shown in the
figure. Similar trends also hold in the non-reciprocal case as shown in Fig.
\ref{fig-PowAllononrecip}, where $\cE_{t0}/\tau_{t0}$ and $\cE_{L1}/\tau_{t0}$  are the round-trip
training powers, $\cE_{L2}/\tau_{L2}$ is the reverse training power, $\cE_{t3}/\tau_{t3}$ is the
forward pilot signal power, and $(\Nt-\NL)\sa2$ is the AN power. Notice that, since the round-trip
and the reverse training work together in this case to provide the transmitter with knowledge of
the downlink channel, both powers should be increased to reduce AN's interference on LR.


\begin{figure}[t]
\begin{center}
\subfigure[Reciprocal case]{\label{fig-NMSE_r39}
\psfrag{Pave}[Bc][Bc]{ $\bPave$ (dB)}
\psfrag{NMSE}[Bc][Bc]{NMSE}
\psfrag{URgamma01}[Bl][Bl]{\small UR,~$\gamma=0.1$}
\psfrag{LRgamma01}[Bl][Bl]{\small LR,~$\gamma=0.1$}
\psfrag{URgamma003}[Bl][Bl]{\small UR,~$\gamma=0.03$}
\psfrag{LRgamma003}[Bl][Bl]{\small LR,~$\gamma=0.03$}
\psfrag{NMSE lower boundbou}[Bl][Bl]{\small NMSE lower bound}

\includegraphics[scale=0.57]{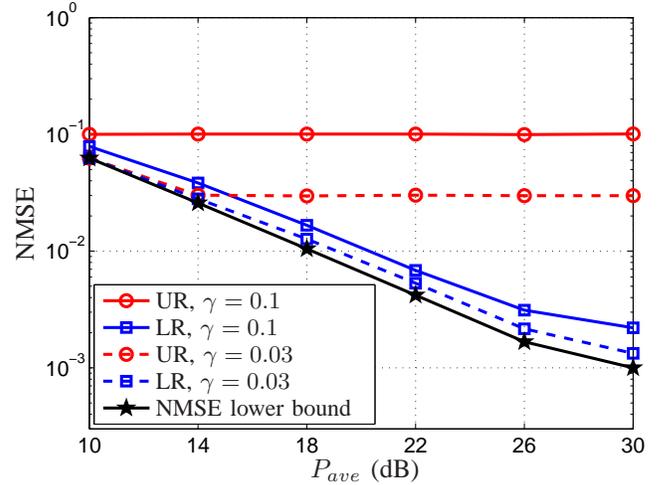}}

\subfigure[Non-reciprocal case]{ \label{fig-NMSE_nonrecip}
\psfrag{Pave}[Bc][Bc]{ $\bPave$ (dB)}
\psfrag{NMSE}[Bc][Bc]{NMSE}
\psfrag{URgamma01}[Bl][Bl]{\small UR,~$\gamma=0.1$}
\psfrag{LRmtgamma01}[Bl][Bl]{\small LR,~Monte Carlo, $\gamma=0.1$}
\psfrag{LRappgamma01}[Bl][Bl]{\small LR,~approx, $\gamma=0.1$}
\psfrag{URgamma003}[Bl][Bl]{\small UR,~$\gamma=0.03$}
\psfrag{LRmtgamma003}[Bl][Bl]{\small LR,~Monte Carlo, $\gamma=0.03$}
\psfrag{LRappgamma003}[Bl][Bl]{\small LR,~approx, $\gamma=0.03$}
\psfrag{NMSE lower boundboundbou}[Bl][Bl]{\small NMSE lower bound}

\includegraphics[scale=0.57]{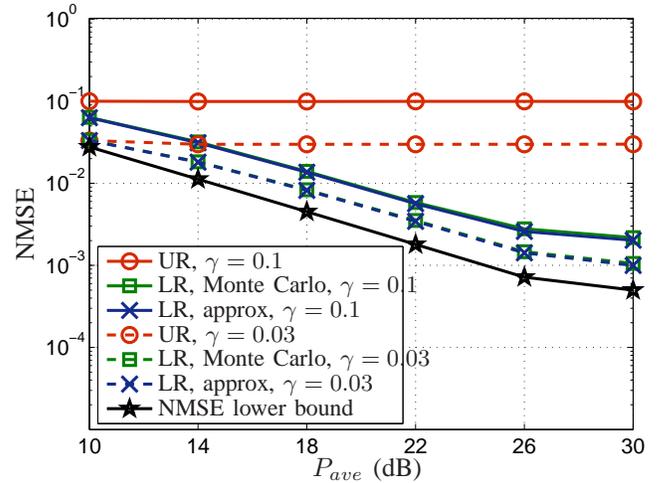}}
\caption{NMSE performance of the proposed DCE schemes for the reciprocal and the non-reciprocal cases.}
\vspace{-0.5cm}
\label{fig-NMSE2}
\end{center}
\end{figure}

In Fig. \ref{fig-NMSE2}, we show the channel estimation performance at LR and UR for different
values of average power budget $P_{ave}$. The reciprocal case is shown in Fig.~\ref{fig-NMSE_r39}
while the non-reciprocal case is in Fig.~\ref{fig-NMSE_nonrecip}. Two different lower limits on the
UR's NMSE are considered in the figures, \ie $\gamma=0.1$ and $\gamma=0.03$. We see that our
proposed DCE schemes can indeed constrain the UR's NMSE above $\gamma$ in both cases. In the case
of non-reciprocal channels, we also compare the approximation of LR's NMSE obtained in
(\ref{eq.NMSEL.appr2}) with the exact value obtained from Monte-Carlo simulations in
Fig.~\ref{fig-NMSE_nonrecip}. We can see that the analytical approximation of the NMSE is very
close to that obtained from Monte-Carlo simulations.

\begin{figure}[t!]
\begin{center}
\subfigure[Reciprocal case: 64-QAM OSTBC]{ \label{fig-SER_rec_64}
\psfrag{Pave}[Bc][Bc]{ $\bPave$ (dB)}
\psfrag{SER}[Bc][Bc]{SER}
\psfrag{URgamma01}[Bl][Bl]{\small UR,~$\gamma=0.1$}
\psfrag{LRgamma01}[Bl][Bl]{\small LR,~$\gamma=0.1$}
\psfrag{URgamma003}[Bl][Bl]{\small UR,~$\gamma=0.03$}
\psfrag{LRgamma003}[Bl][Bl]{\small LR,~$\gamma=0.03$}
\psfrag{Perfect CSI}[Bl][Bl]{\small Perfect CSI}

\includegraphics[scale=0.61]{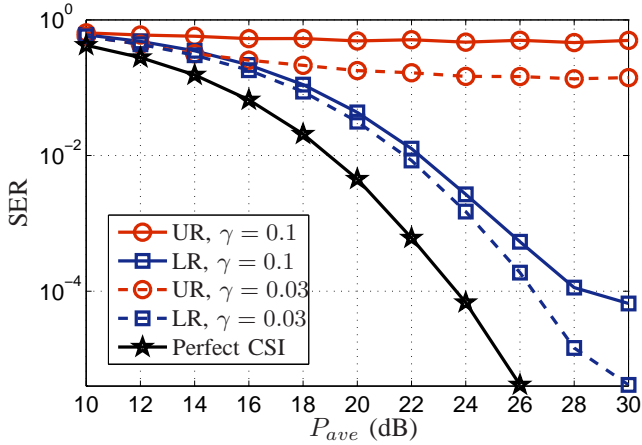}}

\subfigure[Non-reciprocal case: 64-QAM OSTBC]{ \label{fig-SER_nonrecip_64}
\psfrag{Pave}[Bc][Bc]{ $\bPave$ (dB)}
\psfrag{SER}[Bc][Bc]{SER}
\psfrag{URgamma01}[Bl][Bl]{\small UR,~$\gamma=0.1$}
\psfrag{LRgamma01}[Bl][Bl]{\small LR,~$\gamma=0.1$}
\psfrag{URgamma003}[Bl][Bl]{\small UR,~$\gamma=0.03$}
\psfrag{LRgamma003}[Bl][Bl]{\small LR,~$\gamma=0.03$}
\psfrag{Perfect CSI}[Bl][Bl]{\small Perfect CSI}
\includegraphics[scale=0.57]{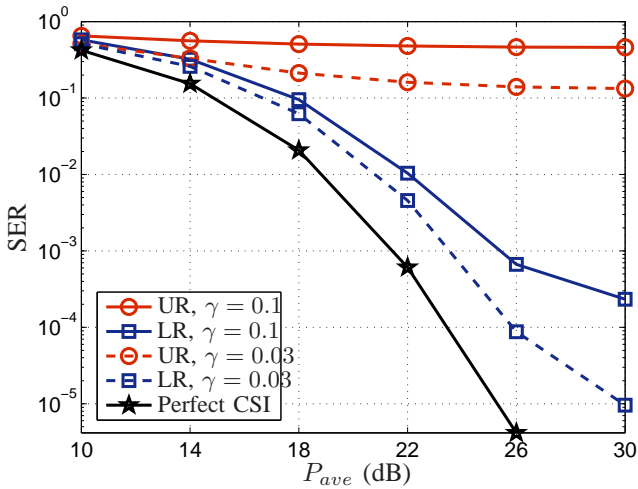}}
\vfill \caption{SER performance of LR and UR in an OSTBC system with channel state information (CSI) obtained by the proposed DCE schemes.}
\label{fig-SER}
\vspace{-0.5cm}
\end{center}
\end{figure}

Finally, in Fig. \ref{fig-SER}, we show the symbol error rate (SER) performance at LR and UR in the
data transmission phase. We consider the scenario where the transmitter sends a $4\times 4$ complex
orthogonal space-time block code (OSTBC). The code length is equal to four and each code block
contains three 64-QAM source symbols~\cite{ref.STBC}. The data transmission power is set equal to
the average transmit power budget $\bPave$. {Note that this $4\times 4$ OSTBC is not
identifiable \cite{Via08,Shap05}, which, as discussed in Remark~1, implies that UR would suffer
from non-trivial rotation ambiguities when either blindly estimating the channel or detecting the
codeword. Therefore, it is assumed here that both LR and UR exploit their channel estimates
obtained under the proposed DCE scheme,} and detect the data symbols by using the coherent
maximum-likelihood detector (assuming that the obtained channel estimates are perfect\footnote{LR
and UR can also employ the CSI-error-aware detector in \cite[Eqn. (20)]{ref.STBC2}, despite that
the associated complexity is much higher especially for higher-order QAM. It is anticipated that
both LR and UR would have improved symbol error performance.})~\cite{ref.STBC}. In this Monte-Carlo
simulation, the SER is computed by averaging over $500,000$ channel realizations and OSTBCs.

Fig.~\ref{fig-SER_rec_64} presents the SER for 64-QAM OSTBC in the reciprocal case. We see that the
SER at LR gradually improves as the average power budget increases, while the SER at UR
remains larger than $0.1$ due to the poor channel estimation. Similar trends are also observed in
the non-reciprocal case in Fig.~\ref{fig-SER_nonrecip_64}. Both figures illustrate that, with the
proposed DCE scheme, discrimination of the data detection performances between LR and UR can be effectively achieved. It is worthwhile to mention that the feedback-and-retraining DCE
scheme proposed in~\cite{ChangChiangHongChi_TSP2010} assumed a perfect feedback channel with no
power consumption and, thus, it is difficult to have a fair performance comparison between the
proposed scheme and that in~\cite{ChangChiangHongChi_TSP2010}.

\section{Conclusions}\label{ch8.conclusion}

In this paper, we have proposed new DCE schemes based on the two-way training methodology for both
reciprocal and non-reciprocal channels. The proposed design drastically decreases the overall
training overhead compared to the original DCE scheme in~\cite{ChangChiangHongChi_TSP2010}.
We obtained analytical results on the MSE of the channel estimation and utilized it to derive the optimal power allocation among pilot signals and AN. The optimal power values in both cases were obtained by minimizing the MSE at LR whilst confining the MSE
at UR above some prescribed value. The presented numerical results have demonstrated the effectiveness of the proposed DCE schemes.

In the current paper, we have derived the DCE scheme based on the LMMSE channel estimator. Since UR
may not be restricted to the use of the LMMSE channel estimator, it would be interesting to extend
the DCE scheme to other more complex channel estimators or even using Cram$\acute{e}$r-Rao lower
bound (CRLB) as the performance measure. Furthermore, by intuition, the channel condition
difference caused by DCE should improve the achievable secrecy rate defined in the context of
information-theoretic security \cite{secrecy.capacity,Khisti,Oggier}. Analytically proving this
intuition, though challenging, is an interesting future research direction. {Interested
readers may refer to
\cite{ChangICC11,Liu2012,LinChangLiangHongChi,Zhou2010,Bloch2009,Rezki2012,Mukherjee2011,Chia2012} for some
endeavors which aim to characterize the impact of channel estimation errors at terminals on the
achievable secrecy rate.}

\section{Acknowledgment}
{The authors would like to sincerely thank the associate editor and the anonymous reviewers whose valuable comments have helped us improve the paper significantly.}

\appendices

\section{Proof of Proposition \ref{prop.optpower.rec}}\label{append.prop1}

{In this appendix, we present the solutions for the following problem
\begin{subequations}\label{opt.prob.I12}
\begin{align}
\min_{\cE_R,\cE_F,\sigma_a^2\geq0}&\mbox{NMSE}_L \\
\mbox{subject to (s.t.)}~~~ &\mbox{NMSE}_U \geq \gamma, \label{opt.prob.I1b2}\\
&\cE_R +\cE_F+(\Nt-\NL)\sigma_a^2\tau_F \leq \bEave,\label{opt.prob.I1c2}\\
&\cE_R \leq \bEL ,\label{opt.prob.I1d2}\\
&\cE_F+(\Nt-\NL)\sigma_a^2\tau_F \leq \bEt, \label{opt.prob.I1e2}
\end{align}
\end{subequations}
where \eqref{opt.prob.I1c2} is a total energy constraint and $\bEave$ denotes the total energy budget.
Note that, when $\bEave >\bEL+\bEt$, the total energy constraint \eqref{opt.prob.I1c2}
is redundant, and thus \eqref{opt.prob.I12} reduces to \eqref{opt.prob.I1}. The solutions of \eqref{opt.prob.I12} are given in the following proposition.
\begin{Prop}\label{prop.optpower.rec2} The solutions to \eqref{opt.prob.I12} with $\gamma$ chosen according to \eqref{cond.gamma}
are given by considering the following three scenarios separately.\\
\noindent\underline{Scenario 1 ($\bEave >\bEL+\bEt$):} For this scenario, problem \eqref{opt.prob.I12} reduces to problem \eqref{opt.prob.I1}.  
The corresponding solution is given in Proposition \ref{prop.optpower.rec}.

\noindent\underline{Scenario 2 ($\max\{\bEL ,\ \bEt \} \leq \bEave \leq \bEL+\bEt$):} If
$$\mu >\min\{\bEL,\ \bEave-\tgam\},$$
the optimal $(\cE_R,\cE_F,\sa2)$ is given by $\cE_R^\star=0$, $\cE_F^\star=\tgam$ and $(\sa2)^\star=0$ (\ie
no need of reverse training and no need of AN in the forward training). On the other hand, if $\mu
\leq \min\{\bEL,\ \bEave-\tgam\}$, the optimal $\cE_R$ can be obtained by solving the following
one-dimensional optimization problem:
\begin{align}\label{opt.prob.final1}
\cE_R^\star=\arg\mathop{\max}_{\cE_R}\ \ \ &\frac{(\NL\swt2+\sigma_H^2\cE_R)\cdot \cE_F(\cE_R)}{\NL\swt2+\sigma_H^2\cdot \cE_R +\NL\sigma_H^2\frac{\sigma_{\tilde{w}}^2}{\sigma_w^2}\cdot \zeta(\cE_R)} \\
\text{\rm s.t.}~~& \max \{0,\mu,\bEave-\bEt \} \leq \cE_R \notag \\
&~~~~~~~~~~~~~~~~~~~~~~~\leq \notag
\min \{\bEL,\ \bEave-\tgam\},
\end{align}
where
\begin{equation}
\zeta(\cE_R)=\frac{\bEave-\tgam-\cE_R}{\tau_F+\sigma_g^2\tgam/\sigma_v^2}
\end{equation}
and
\begin{equation}
\cE_F(\cE_R)=\tgam\left(\frac{\sigma_g^2}{\sigma_v^2}\cdot \zeta(\cE_R)+1\right).
\end{equation} The optimal $\cE_F$ and $\sigma_a^2$ are given by
$\cE_F^\star=\cE_F(\cE_R^\star)$ and $(\sa2)^\star=\frac{\zeta(\cE_R^\star)}{(\Nt-\NL)}$.

\noindent\underline{Scenario 3 ($\bEave < \bEL$ and/or $\bEave < \bEt$):} This scenario refers to the case where one or both
individual energy constraint(s) are redundant. Here, the solution is given as in Scenario $2$ with the redundant individual energy constraint(s) set to infinity.
\end{Prop}}

We first prove the most general scenario, i.e., Scenario 2, where both the individual and total power constraints are effective. Scenario 1 and scenario 3 are degenerated cases of Scenario 2; their corresponding solutions are presented in the second subsection.

\subsection{Proof of Scenario 2}
For
notational simplicity, let us define $\zeta=(\Nt-\NL)\sa2$. In this case, Problem \eqref{opt.prob.I1} can be rewritten as
\begin{subequations}\label{opt.prob.II.reformulation}
\begin{align}
\!\!\!\!\!\!\max_{\cE_F,\cE_R,\zeta\geq 0}\ \ &
\frac{1}{\sH2}+\frac{1}{\Nt\sigma_w^2}\cdot\frac{(\NL\sigma_{\tilde
w}^2+\sH2\cE_R)\cE_F}{\NL\swt2+\sH2 \cE_R
+\NL\sH2\frac{\sigma_{\tilde{w}}^2}{\sigma_w^2} \zeta} \\
\mbox{s.t. }~ &\frac{\sigma_v^2 \cE_F}{\sG2
\zeta+\sigma_v^2}\leq \tgam,\\
&\cE_R + \cE_F+\zeta\cdot \tau_F  \leq \bEave, \label{opt.prob.II.reformulationC3}\\
&\cE_R\leq\bEL\\
&\cE_F+\zeta\cdot \tau_F\leq \bEt.
\end{align}
\end{subequations}
We will analyze the solutions of \eqref{opt.prob.II.reformulation} via the following two steps: (i)
for any given $\cE_R$, find the optimal values of $\cE_F$ and $\zeta$ as functions of $\cE_R$; and (ii) find the optimal
value of $\cE_R$.

\noindent\underline{\bf Step (i)}:
Given $\cE_R$, where $0\leq \cE_R\leq \min\{\bEL,\bEave\}$, the optimal values of $\cE_F$ and $\zeta$ can be found equivalently by solving the following optimization problem:
\begin{subequations}\label{opt.prob.II.givenER}
\begin{align}\label{opt.prob.II1}
\max_{\cE_F,\zeta\geq 0}\ \
&\frac{(\NL\sigma_{\tilde w}^2+\sH2\cE_R)\cE_F}{\NL\swt2+\sH2 \cE_R
+\NL\sH2\frac{\sigma_{\tilde{w}}^2}{\sigma_w^2}\cdot \zeta} \\
\mbox{s.t.}~ &\frac{\sigma_v^2 \cE_F}{\sG2
\zeta+\sigma_v^2}\leq \tgam,\label{opt.prob.II2}\\
&\cE_F+\zeta\cdot \tau_F \leq \bEave-\cE_R,\\
&\cE_F+\zeta\cdot \tau_F\leq \bEt \label{opt.prob.II4}.
\end{align}
\end{subequations}
Let the solutions to $\cE_F$ and $\zeta$ in the above problem be denoted by $\cE_F^\star(\cE_R)$ and $\zeta^\star(\cE_R)$, respectively. To analyze \eqref{opt.prob.II.givenER}, we consider the following two cases:

\noindent\underline{Case 1 ($\bEave-\tgam<\cE_R \leq \min\{\bEL,\bEave\}$):} 
Note that the objective function in \eqref{opt.prob.II.givenER} is monotonically increasing in
$\cE_F$ but decreasing in $\zeta$. Since $\tgam$ satisfies \eqref{cond.tigamma} and $\cE_R>\bEave-\tgam$, it follows that $\zeta^\star(\cE_R)=0$ and $\cE_F^\star(\cE_R)=\bEave-\cE_R$. {It can be easily verified that $\zeta^\star(\cE_R)$ and $\cE_F^\star(\cE_R)$ are feasible to \eqref{opt.prob.II.givenER}. In particular, \eqref{opt.prob.II4} is satisfied due to $\cE_F^\star(\cE_R)=\bEave-\cE_R$ and $\tgam \leq \bEt$ by \eqref{cond.tigamma}.}
In this case, the maximum objective value of \eqref{opt.prob.II1} is equal to
$\bEave-\cE_R$,
which is less than $\tgam$ {due to the premise of this case}.

\noindent\underline{Case 2 ($\cE_R\leq \min\{\bEL,\ \bEave-\tgam \}$):} In this case, we first show that
the constraint in (\ref{opt.prob.II2}) must
hold with equality when the optimum value is achieved.
Suppose that
the constraint in (\ref{opt.prob.II2}) is inactive at the optimum. In this case,  we can always decrease $\zeta$ to
obtain a larger objective value until the constraint (\ref{opt.prob.II2}) holds with equality. If
the condition (\ref{opt.prob.II2}) is still inactive even when $\zeta=0$, we can instead lift
$\cE_F$ to achieve a larger objective value since
$\cE_R\leq \min\{\bEL,\ \bEave-\tgam \}$ and $\tgam$ must satisfy \eqref{cond.tigamma}.
Hence, we conclude that the constraint in (\ref{opt.prob.II2}) must
hold with equality and, thus,
\begin{equation} \label{optbc.equofa}
\cE_F^\star(\cE_R)=\tgam\left( \frac{\sigma_g^2}{\sigma_v^2}\cdot \zeta^\star(\cE_R)+1\right).
\end{equation}
By substituting (\ref{optbc.equofa}) into \eqref{opt.prob.II.givenER},
the optimization problem can be reduced to
\begin{subequations}\label{opt.prob.III.givenER}
\begin{align}\label{opt.prob.III1}
\max_{\zeta\geq 0}~~~ &\frac{({\sG2}/{\sigma_v^2}\cdot
\zeta+1)(\NL\swt2+\sH2
\cE_R)\tgam}{\NL\sH2\frac{\sigma_{\tilde{w}}^2}{\sigma_w^2}\zeta+\NL\swt2+\sH2 \cE_R} \\
\mbox{s.t.}\ \ \ \ &\left( \tau_F+\frac{\sG2\tgam}{\sigma_v^2}\right)\zeta \leq
\bEave-\cE_R-\tgam,
\label{opt.prob.III2}\\
& \left( \tau_F+\frac{\sG2\tgam}{\sigma_v^2}\right)\zeta \leq
\bEt-\tgam.\label{opt.prob.III3}
\end{align}\end{subequations}
We further consider the following two subranges of $\cE_R$ in order to find the optimal $\zeta$. Let us iterate the definition of $\mu$ in \eqref{condition u}
\begin{align} \notag
\mu \triangleq \NL
\left(\frac{\sigma_v^2\sigma_{\tilde{w}}^2}{\sigma_g^2\sigma_w^2}
-\frac{\sigma_{\tilde{w}}^2}{\sigma_h^2}\right)
>\min\{\bEL,\ \bEave-\tgam\}.
\end{align}
\begin{enumerate}
\item[(a)] \underline{$\cE_R < \mu$}:
It can be shown that the objective function in (\ref{opt.prob.III1}) is monotonically decreasing in
$\zeta$ whenever $\cE_R < \mu$. Therefore, for $\cE_R < \mu$, the optimal $\zeta$ of
\eqref{opt.prob.III.givenER} is zero and the corresponding optimal objective value is given by
$\tgam$.

\item[(b)] \underline{$\mu \leq \cE_R$}: On the other hand, when $\mu \leq \cE_R$,
the objective function in (\ref{opt.prob.III1}) is
monotonically non-decreasing in $\zeta$. Hence, if $\cE_R \leq \bEave-\bEt$,
one can increase $\zeta$ until constraint (\ref{opt.prob.III3}) is met with equality. In this case, we have
\begin{equation}\label{equ.c_a1}
\zeta^\star(\cE_R)=\frac{\bEt-\tgam}{\tau_F+\sG2\tgam/\sigma_v^2}.
\end{equation}
Conversely, if $\cE_R\geq \max\{\mu,\
\bEave-\bEt\}$,
constraint (\ref{opt.prob.III2}) must hold with equality at the optimum and, thus,
\begin{align}\label{equ.c_a}
\zeta^\star(\cE_R)=\frac{\bEave-\cE_R-\tgam} {\tau_F+\sG2\tgam/\sigma_v^2}.
\end{align}
Moreover, since $\cE_R\geq \mu$ implies that $\sG2/\sigma_v^2 \geq \frac{\NL\sH2\sigma_{\tilde{w}}^2/\sigma_w^2}{\NL\swt2+\sH2
\cE_R}$, the optimal objective value of
\eqref{opt.prob.III1}, \ie
\begin{align}\label{obj.inequ.agmu}
\frac{\sG2/\sigma_v^2
\zeta^\star+1}{\frac{\NL\sH2\sigma_{\tilde{w}}^2/\sigma_w^2}{\NL\swt2+\sH2
\cE_R}\zeta^\star+1}\cdot \tgam
\end{align}
is no less than $\tgam$.
\end{enumerate}

Notice that the maximum objective value obtained in Case 2 is no less than $\tgam$ and, thus, is always greater than that obtained in Case 1.

\noindent\underline{\bf Step (ii) :}
Since the maximum objective value in Case 2 is always greater than that in Case 1, the optimal value of $\cE_R$ must satisfy $\cE_R\leq\min\{\bEL,\ \bEave-\tgam\}$.

If $\mu >\min\{\bEL,\ \bEave-\tgam\}$, then it follows that $\cE_R<\mu$ since  $\cE_R\leq\min\{\bEL,\ \bEave-\tgam\}$. Therefore, by the results of Case 2(a), we have $\zeta^\star=0$ and, thus, $\cE_F^\star=\tgam$ by
\eqref{optbc.equofa}. Since AN is not needed, there is also no need for reverse training and, thus, we can set $\cE_R^\star=0$.
%
Alternatively, if $\mu \leq \min\{\bEL,\
\bEave-\tgam\}$, $\cE_R$ can be chosen to be either greater or smaller than $\mu$. However, by the results of {Case 2}, we know that a larger objective can be achieved when $\cE_R\geq \mu$. Therefore,
the optimal $\cE_R$ must lie in the range
$\max\{0,\ \mu\}\leq \cE_R
\leq \min\{\bEL,\ \bEave-\tgam\}$.

Let us first consider the subrange  $\max\{0,\mu\}
\leq \cE_R \leq \bEave-\bEt$ $(\leq \bEL)$, if it exists. In this case, the optimal values of $\cE_F$ and $\zeta$ are given by
\eqref{optbc.equofa} and
\eqref{equ.c_a1}, which actually do not depend on $\cE_R$. By replacing $\cE_F$ and $\zeta$ with their optimal values $\cE_R(\cE_F)$ and $\zeta(\cE_F)$, the original problem \eqref{opt.prob.II.reformulation} can be written as
\begin{align} \label{opt.prob.sub1}
\max_{{\cE}_R \geq 0}\ \ \
&\frac{(\NL\swt2+\sH2{\cE}_R)\cE_F^\star({\cE}_R)}{\NL\swt2+\sH2 {\cE}_R +\NL\sH2\frac{\sigma_{\tilde{w}}^2}{\sigma_w^2} \zeta^\star({\cE}_R)} \\
\mbox{s.t.} \ \ \ &\max\{0,\ \mu\}\leq {\cE}_R \leq \bEave-\bEt. \notag
\end{align}
Notice that, since $\cE_F^\star({\cE}_R)$ and $\zeta^\star({\cE}_R)$ in
\eqref{optbc.equofa} and \eqref{equ.c_a1} do not depend on $\cE_R$, the objective function (\ref{opt.prob.sub1}) is monotonically non-decreasing
with respect to ${\cE_R}$ and, thus, the optimal value is
achieved with ${\cE}_R=\bEave-\bEt$. Therefore, it is sufficient to consider $\cE_R$ in the subrange
$\max \{0,\ \mu,\ \bEave-\bEt\} \leq \cE_R
\leq \min \{\bEL,\ \bEave-\tgam\}$ since it includes the value ${\cE}_R=\bEave-\bEt$. It then follows from (\ref{equ.c_a}) and
(\ref{optbc.equofa}) that the optimal ${\cE_R}$ can be obtained by solving (\ref{opt.prob.final1}), which requires only a simple line
search over the finite interval.
%
%
%

{\subsection{Proof of Scenario 1 and Scenario 3}}
In Scenario 1, where $\bEave >\bEL+\bEt$, the total energy constraint
in \eqref{opt.prob.II.reformulationC3} is redundant and, thus,
$\bEave$ can be set as infinity. In particular, if $\mu>\bar \cE_L$ (and thus $\mu>\cE_R$), we have $\zeta^\star=0$, $\cE_F^\star=\tgam$
and $\cE_R^\star=0$ according to Case 2(a). On the other hand, if $\mu\leq \bar \cE_L$, it follows from Case 2(b) that
the original problem can be expressed as
\begin{align} \label{opt.prob.sub99}
\max_{{\cE}_R\geq 0}\ \ \
&\frac{(\NL\sigma_w^2+\sH2{\cE}_R)\cE_F^\star({\cE}_R)}{\NL\sigma_w^2+\sH2\cdot {\cE}_R +\NL\sH2\frac{\sigma_{\tilde{w}}^2}{\sigma_w^2}\cdot \zeta^\star({\cE}_R)} \\
\mbox{s.t}\ \ \ \ \ &\max\{0,\ \mu\}\leq {\cE}_R \leq \bEL, \notag
\end{align}
where $\cE_F^\star(\cE_R)$ and $\zeta^\star(\cE_R)$ are given by \eqref{optbc.equofa} and
\eqref{equ.c_a1}. Since $\cE_F^\star(\cE_R)$ and $\zeta^\star(\cE_R)$ do not depend on $\cE_R$ in this case, the objective in \eqref{opt.prob.sub99} increases monotonically with $\cE_R$ and, thus, the optimal value of $\cE_R$ is given by $\cE_R^\star=\bEL$. This implies that both LR and transmitter should transmit with their maximum energies in Scenario 3.

In Scenario $3$, where $\bEave < \bEL$ and/or $\bEave < \bEt$, at least one of the individual energy constraints are redundant and, thus, can be set as infinity. Therefore, the optimal solution can be obtained similarly by solving (\ref{opt.prob.final1}) with the redundant constraint(s) (\ie $\bar \cE_L$ and/or $ \bEt$) set as infinity.


\section{Proof of the Optimal Pilot Matrix $\bC_t$}\label{append.optbCt}

Consider the optimization of $\bC_t$ for problem \eqref{opt.prob.I1}. Notice, from \eqref{NMSE2.forward.channel.LR} and \eqref{NMSE2.forward.channel.UR}, that the NMSE at both receivers depend only on the value of $\bC_t^H\bC_t$. Let
\begin{equation}\label{evd}
\bC^H_t\bC_t=\mathbf{U}_c\mathbf{D}\mathbf{U}_c^H
\end{equation}
be the eigenvalue
decomposition of $\bC^H_t\bC_t$,
where $\mathbf{U}_c\in\mathbb{C}^{\Nt\times \Nt}$ is a unitary matrix and $\mathbf{D}=\diag(d_1,\dots,d_{\Nt})$ is a diagonal matrix with $d_1,\dots,d_{\Nt}$ being
the eigenvalues of $\bC^H\bC$.
By substituting \eqref{evd} into \eqref{NMSE2.forward.channel.LR} and \eqref{NMSE2.forward.channel.UR}, we have $$\mbox{NMSE}_L=\frac{1}{\Nt}\overset{\Nt}{\underset{i=1}{\sum}}
\left(\frac{1}{\sH2}+a\cdot d_i\right)^{-1}$$ and $$\mbox{NMSE}_U=\frac{1}{\Nt}\overset{\Nt}{\underset{i=1}{\sum}}
\left(\frac{1}{\sG2}+b\cdot d_i\right)^{-1},$$ where $$a=\frac{\cE_F/\Nt}
{(\Nt-\NL)\left(\frac{1}{\sH2}
+\frac{\cE_R}{\NL\sigma_{\tilde{w}}^2}\right)^{-1}\sigma_a^2+\sigma_w^2}$$ and
$$b=\frac{\cE_F/\Nt}
{(\Nt-\NL)\sG2\sigma_a^2+\sigma_v^2}.$$
Therefore, given $\cE_R$, $\cE_F$, and $\sigma_a^2$,  the optimal $\bC_t$ can be found by solving
the following optimization problem
\begin{subequations}\label{CC}
\begin{align}
\min_{d_1,\dots,d_{\Nt}\geq 0}&\frac{1}{\Nt}\overset{\Nt}{\underset{i=1}{\sum}}
\left(\frac{1}{\sH2}+a\cdot d_i\right)^{-1} \\
\mbox{s.t.}~~~&\frac{1}{\Nt}\overset{\Nt}{\underset{i=1}{\sum}}
\left(\frac{1}{\sG2}+b\cdot d_i\right)^{-1}\geq \gamma \label{CC2}\\
&\overset{\Nt}{\underset{i=1}{\sum}} d_i=\Nt,\ d_i\geq0\ \mbox{for}\ i=1,\dots,\Nt \label{CC1}
\end{align}
\end{subequations}
where \eqref{CC1} is due to the constraint that ${\rm Tr}(\bC^H_t\bC_t)=\Nt$. By the Karush-Kuhn-Tucker (KKT) conditions, the optimal $d_i$ must satisfy the following conditions
\begin{align}
&-\frac{a}{\Nt}\left(\frac{1}{\sH2}+a d_i\right)^{-1}
+\frac{\kappa b}{\Nt}\left(\frac{1}{\sG2}+b d_i\right)^{-1}
 \notag \\
&~~~~~~~~~~~~~~~~~~~~~~~~~~~+\nu-\eta_i=0,~i=1,\dots,\Nt \label{equ.KKT1}\\
&\overset{\Nt}{\underset{i=1}{\sum}} d_i=\Nt,\ \kappa \geq 0,\ \nu\geq0, \label{equ.KKT2} \\
&\eta_id_i=0,\ \eta_i\geq 0,\ d_i\geq 0,\ i=1,\dots,\Nt, \label{equ.KKT3}
\end{align}
where  $\kappa$, $\nu$ and $\eta_i$ are the corresponding dual variables of the constraints in \eqref{CC2} and \eqref{CC1}. It follows from \eqref{equ.KKT3} that $\eta_i=0$ if $d_i>0$, and therefore, by \eqref{equ.KKT1}, we can observe that all the nonzero $d_i$ must have the same value. Hence, if there are  $K$ nonzero $d_i$'s, then, owing to \eqref{equ.KKT2}, we have $d_1=\cdots=d_K=\Nt/K$.

\section{Derivation of LR's NMSE in Non-Reciprocal Case}
\label{append.deriv.nonNMSEL}

Here we derive the downlink channel estimation performance at the LR. By expressing $\bHd=\widehat\bH_d-\Delta\bHd$ and the fact that $\bNHdHt\hbHdt=\mathbf{0}$,
the received signal at LR in (\ref{eq.receive.signal.LR3}) can be written as
\begin{equation}\label{eq.receive.signal.LR2}
\bY_{L3}=\bar{\bC}_{t3}\bHd-\bA\bNHdHt\Delta\mathbf{H}_{d,t}+\bW_3.
\end{equation}
where $\bar{\bC}_{t3}\triangleq\sqrt{\frac{\cE_{t3}}{\Nt}}\bC_{t3}$. Its vector representation is given by \begin{equation}\label{eq.receive.signal.LR22}
\mathbf{y}_{L3}=\left(\bI_{\NL}\otimes \bar{\bC}_{t3}\right)\mathbf{h}_d
-(\bI_{\NL}\otimes \bA\bNHdHt)\Delta\mathbf{h}_{d,t}+\mathbf{w}_3
\end{equation}where $\mathbf{y}_{L3}=\text{vec}(\bY_{L3})$ and $\mathbf{w}_3=\text{vec}(\bW_3)$. The LMMSE estimate of $\mathbf{h}_d$ is given by
\begin{equation}
\mathbf{\hat{h}}_d=\bR_{\bh_d \by_{L3}}\bR^{-1}_{\by_{L3}\by_{L3}}\mathbf{y}_{L3}
\end{equation}
where
\begin{align}\label{eq.covar.hdyL3}
\bR_{\bh_d\by_{L3}}=\mathbb{E}\{\mathbf{h}_d\mathbf{y}_{L3}^H\}=\sHd2\left(\bI_{\NL}\otimes \bar{\bC}_{t3}\right)
\end{align}
and
\begin{align}
&\bR_{\by_{L3}\by_{L3}}=\mathbb{E}\{\mathbf{y}_{L3}\mathbf{y}_{L3}^H\}=\sHd2\left(\bI_{\NL}\otimes \bar{\bC}_{t3}\bar{\bC}_{t3}^H\right) \notag\\
&\label{eq.covar.yL3}
+\mathbb{E}\{(\bI_{\NL}\otimes \bA\bNHdHt)\Delta\mathbf{h}_{d,t}
\Delta\mathbf{h}^H_{d,t}(\bI_{\NL}\otimes \bA\bNHdHt)^H\}
\notag\\
&+\sw2(\bI_{\NL}\otimes \bI_{\Nt}).
\end{align}
Note that the expectation in (\ref{eq.covar.yL3})
is taken over all the random variables including $\bA$, $\Delta\mathbf{h}_{d,t}$ and
$\hbHdt$, where the last two variables actually depend on the value of $\hbHu$.
Using the law of iterated expectations, \ie $\mathbb{E}\{X\}=\mathbb{E}\{\mathbb{E}\{X|Y\}\}$,
the second term of (\ref{eq.covar.yL3}) can be written as
\begin{align}\label{eq.covar.yL3.2term}
&\mathbb{E}_{\hbHu}\{ \mathbb{E}_{\bA,\hbHdt}\{(\bI_{\NL}\otimes \bA\bNHdHt)
\notag \\
&\times \mathbb{E}\{\Delta\mathbf{h}_{d,t}\Delta\mathbf{h}^H_{d,t}|\hbHdt,\hbHu\}
(\bI_{\NL}\otimes \bA\bNHdHt)^H|\hbHu  \}\}
\end{align}
where we have used the fact that the random matrix $\bA$ is independent of $\Delta\mathbf{h}_{d,t}$. Since $\Delta\mathbf{h}_{d,t}$ and $\hbHdt$ are not necessarily independent, it is difficult to evaluate $\mathbb{E}\{\Delta\mathbf{h}_{d,t}\Delta\mathbf{h}^H_{d,t}|\hbHdt,\hbHu\}$. To obtain a tractable form, we consider an approximation where
$\Delta \mathbf{h}_{d,t}$ and $\hbHdt$ are assumed to be independent, \ie $\mathbb{E}\{\Delta\mathbf{h}_{d,t}\Delta\mathbf{h}^H_{d,t}|\hbHdt,\hbHu\}\approx \mathbb{E}\{\Delta\mathbf{h}_{d,t}\Delta\mathbf{h}^H_{d,t}|\hbHu\}$.
By (\ref{eq.mse.downlinkHd.tx}) and the fact that $\bNHdHt\bNHdt=\bI_{\Nt-\NL}$, equation~\eqref{eq.covar.yL3.2term} can be computed as
\begin{align}\label{eq.covar.yL3.2term.3}
&(\Nt-\NL)\sa2\bigg[\sHd2\bI_{\NL}-\sHd2\frac{\sHd2 \cE_{t0}}{\sHd2 \cE_{t0}+\Nt\sw2}\notag \\
&\times\mathbb{E}_{\hbHu}\Biggl\{\bigg(
\bigg(\frac{1}{\beta}\hbHu^*\hbHu^T\bigg)^{-1}+\bI_{\NL}\bigg)^{-1}\Biggr\}\bigg]\otimes \bI_{\Nt},
\end{align}
where $\beta$ is defined in \eqref{eq.beta}. To further evaluate \eqref{eq.covar.yL3.2term.3}, let us take the eigenvalue decomposition of $\hbHu\hbHu^H$ as
$\mathbf{U}\mathbf{\Lambda}\mathbf{U}^H$,
where $\mathbf{U}\in \bbC^{\NL\times \NL}$ is a unitary matrix and $\mathbf{\Lambda}=\diag(\lambda_1,\dots,\lambda_{\NL})$
is a diagonal matrix containing the unordered eigenvalues
of $\hbHu\hbHu^H$ as the diagonal elements. It is not difficult to show that the coefficients of $\hbHu$ are \iid Gaussian distributed because both the uplink channel $\bHu$ and the noise
matrix $\bWt_2$ are \iid Gaussian distributed and that the pilot matrix $\sqrt{\frac{\cE_{L2}}{\NL}}\bC_{L2}$
in the reverse training stage (see Section \ref{ch4.NonRecip-A}) is semi-unitary. It then follows from results in random matrix theory \cite{Matrix_variable_dist} that $\hbHu\hbHu^H$ has a Wishart
distribution with $\Nt$ degrees of freedom and its mean is given by
\begin{equation}
\mathbb{E}\{\hbHu\hbHu^H\}=\Nt \frac{\sigma_{H_u}^4 \cE_{L2}}{\sHu2 \cE_{L2}+\NL\swt2}\bI_{\NL}
\triangleq\Nt \sigma^2\bI_{\NL}\notag
\end{equation}
where
$\sigma^2$ is as defined in \eqref{eq.sigma}. 
Since $\mathbf{\Lambda}$ and $\mathbf{U}$ are statistically independent \cite{Random_matrix_theo}, (\ref{eq.covar.yL3.2term.3}) can be further evaluated as
\begin{align} 
&(\Nt-\NL)\sa2\bigg[\sHd2\bI_{\NL}-\sHd2\frac{\sHd2 \cE_{t0}}{\sHd2 \cE_{t0}+\Nt\sw2}
\notag \\
&~~~\times\mathbb{E}_{\mathbf{U}}\Biggl\{\mathbf{U}\cdot
\mathbb{E}_{\mathbf{\Lambda}}\biggl\{\bigg(\beta\mathbf{\Lambda}^{-1}+\bI_{\NL}\bigg)^{-1}\biggr\}
\mathbf{U}^H\Biggr\}\bigg]\otimes \bI_{\Nt} \notag \\
&=(\Nt-\NL)\sa2\bigg[\sHd2-\sHd2\frac{\sHd2 \cE_{t0}}{\sHd2 \cE_{t0}+\Nt\sw2}\label{eq.covar.yL3.2term.5}
\notag \\
&~~~\times\mathbb{E}_{\lambda_1}\biggl\{\bigg(\frac{1}{\beta/\lambda_1+1}\bigg)\biggr\}
\bigg]\bI_{\NL}\otimes \bI_{\Nt}
\end{align}
where the equality follows from the fact that the eigenvalues of the Wishart distributed matrix $\hbHu\hbHu^H$
are identically distributed \cite{Unordered_eigenvalue_dist}.
Substituting (\ref{eq.covar.yL3.2term.5}) into (\ref{eq.covar.yL3}),
we have an approximation of the covariance matrix of $\mathbf{y}_{L3}$ as
\begin{align}
\nonumber&\bR_{\by_{L3}\by_{L3}}\approx\\
& \bI_{\NL}\!\otimes\! \Biggl\{\!\sHd2\bar{\bC}_{t3}\bar{\bC}_{t3}^H
+\bigg[(\Nt-\NL)\sa2\bigg(\sHd2-\sHd2\notag \\
&\times\frac{\sHd2 \cE_{t0}}{\sHd2 \cE_{t0}+\Nt\sw2}
\mathbb{E}_{\lambda_1}\biggl\{\bigg(\frac{1}{\beta/\lambda_1+1}\bigg)\biggr\}\bigg)+\sw2\bigg]\bI_{\Nt}
\!\Biggr\}\label{eq.covar.yL3.2}
\end{align}
Since the NMSE of $\hbHd$ is
\begin{align}
\!\!\!\text{NMSE}_L&=\frac{\text{Tr}(\mathbb{E}\{\Delta\mathbf{h}_d\Delta\mathbf{h}^H_d\})}{\Nt\NL}
\notag \\
&
=\frac{\text{Tr}\left(\sHd2\bI_{\NL\Nt}-\bR_{\bh_d\by_{L3}}\bR^{-1}_{\by_{L3}\by_{L3}}\bR^H_{\bh_d\by_{L3}}\right)}{\Nt\NL},
\label{eq.NMSEL.def}
\end{align}
by substituting (\ref{eq.covar.hdyL3}) and (\ref{eq.covar.yL3.2}) into (\ref{eq.NMSEL.def}), we obtain
an approximation of $\text{NMSE}_L$ as shown in \eqref{eq.NMSEL.appr1} (top of the next page).
\begin{figure*}[t]
\begin{equation}\label{eq.NMSEL.appr1}
\text{NMSE}_L\approx \frac{1}{\Nt}
\mbox{Tr}\left(\frac{1}{\sHd2}\bI_{\Nt}+\frac{\cE_{t3}}{\Nt}\frac{\bC^H_{t3}\bC_{t3}}
{(\Nt-\NL)\sa2\left(\sHd2-\sHd2\frac{\sHd2 \cE_{t0}}{\sHd2 \cE_{t0}+\sw2}
\mathbb{E}_{\lambda_1}\biggl\{\left(\frac{1}{\beta/\lambda_1+1}\right)\biggr\}\right)+\sw2} \right)^{-1}.
\end{equation}
\hrulefill\vspace{.0cm}
\end{figure*}
To further obtain an approximation for the expectation term in
(\ref{eq.NMSEL.appr1}), we note that, when $\Nt\gg 1$, the distribution of the eigenvalues of $\hbHu\hbHu^H$ can be asymptotically
approximated by a Gaussian random variable \cite{Limiting_dist_eigenvalue}, that is $\lambda_1\overset{a.}{\thicksim} \mathcal{N}(\Nt\sigma^2,\Nt\sigma^4)$. 
{Moreover, when $N_L$ is also sufficiently large, $\sigma^2$ in \eqref{eq.sigma} is close to zero (i.e., $\lambda_1$ is approximately a constant equal to its mean), and thus
the term $\mathbb{E}_{\lambda_1}\biggl\{\left(\frac{1}{\beta/\lambda_1+1}\right)\biggr\}$ can be approximated by $\frac{1}{\beta/\mathbb{E}_{\lambda_1}\{\lambda_1\}+1}$
by the Jensen's
inequality. As a result, we obtain \eqref{eq.NMSEL.appr2} as an approximation of \eqref{eq.NMSEL.appr1}.}


\end{document}